\documentclass[10pt,journal,compsoc]{IEEEtran}
\usepackage{algorithm,algorithmicx,amsbsy,amsmath,amssymb,epsfig,bbm,mathrsfs,multirow,amsthm,setspace,textcomp,xcolor,url}
\ifCLASSOPTIONcompsoc
\usepackage[nocompress]{cite}
\else
\usepackage{cite}
\fi

\DeclareMathOperator*{\argmax}{argmax}
\usepackage{subcaption}
\usepackage[labelformat=simple]{subcaption}

\usepackage{eurosym}

\newtheorem{theorem}{Theorem}
\newtheorem{lemma}{Lemma}
\DeclareMathAlphabet\mathbfcal{OMS}{cmsy}{b}{n}
\usepackage{array}

\begin{document}
	
	\title{BlockRoam: Blockchain-based Roaming Management System for Future Mobile Networks}
	
\author{Cong T. Nguyen, Diep N. Nguyen, Dinh Thai Hoang, Hoang-Anh Pham,\\ Nguyen Huynh Tuong, Yong Xiao, and Eryk Dutkiewicz%
	\IEEEcompsocitemizethanks{	\IEEEcompsocthanksitem Cong T. Nguyen, Hoang-Anh Pham, and Nguyen Huynh Tuong are with the Ho Chi Minh City University of Technology, VNU-HCM, Vietnam. E-mail:  \{ntcong.sdh19, anhpham, htnguyen\}@hcmut.edu.vn.
	
	\IEEEcompsocthanksitem Diep N. Nguyen, Dinh Thai Hoang, and Eryk Dutkiewicz are with the School of Electrical and Data Engineering, University of Technology Sydney, Australia.	E-mail:  \{diep.nguyen, hoang.dinh, eryk.dutkiewicz\}@uts.edu.au.

	\IEEEcompsocthanksitem Yong Xiao is with the School of Electronic Information and Communications, Huazhong University of Science and Technology, Wuhan, China. E-mail: yongxiao@hust.edu.cn
	}
    
	\thanks{}
		\vspace{-5mm}}

\IEEEtitleabstractindextext{	
	\begin{abstract}
		Mobile service providers (MSPs) are particularly vulnerable to roaming frauds, especially ones that exploit the long delay in the data exchange process of the contemporary roaming management systems, causing multi-billion dollars loss each year. In this paper, we introduce BlockRoam, a novel blockchain-based roaming management system that provides an efficient data exchange platform among MSPs and mobile subscribers. Utilizing the Proof-of-Stake (PoS) consensus mechanism and smart contracts, BlockRoam can significantly shorten the information exchanging delay, thereby addressing the roaming fraud problems. Through intensive analysis, we show that the security and performance of such PoS-based blockchain network can be further enhanced by incentivizing more users (e.g., subscribers) to participate in the network. Moreover, users in such networks often join stake pools (e.g., formed by MSPs) to increase their profits. Therefore, we develop an economic model based on Stackelberg game to jointly maximize the profits of the network users and the stake pool, thereby encouraging user participation. We also propose an effective method to guarantee the uniqueness of this game's equilibrium. The performance evaluations show that the proposed economic model helps the MSPs to earn additional profits, attracts more investment to the blockchain network, and enhances the network's security and performance.
	\end{abstract}
	
	}
	
		\maketitle
	\IEEEdisplaynontitleabstractindextext
	
	\IEEEpeerreviewmaketitle
	
	%=============================================================
	%=============================================================
	\IEEEraisesectionheading{\section{Introduction}}
	\label{sec:introduction}
	\subsection{Motivation}
	\IEEEPARstart{W}{ith} the popularity of IT technologies and smart devices, over 5 billion people have been subscribed to mobile services, generating a \$1.03 trillion revenue globally in 2018~\cite{revenue}. Although the number of subscribers and the revenues will continue to grow, mobile service providers (MSPs) have been facing several obstacles, especially for roaming services. Among them, fraud management is one of the biggest challenges for MSPs with over \$32.7 billion annual loss throughout the world~\cite{loss}. Roaming fraud exploits the inefficiency in managing data exchanges between two MSPs in order to use illegal free-riding services. In particular, when a subscriber moves from its Home Public Mobile Network (HPMN) to a Visited Public Mobile Network (VPMN) and remotely accesses services of the HPMN via the VPMN's facilities, the HPMN has to pay the VPMN for the subscriber's service usage costs incurred according to the roaming agreement. However, the HPMN may not be able to charge the subscriber properly due to the delay in data exchange between the HPMN and VPMN, i.e., the time interval between when the subscriber finished using the service and when the HPMN received the service report from the VPMN. For example, a subscriber can fraudulently obtain subscription from the HPMN, e.g., by SIM cloning, and uses roaming services in the VPMN. Such roaming fraud can only be detected and responded to after the HPMN receives the service report, which might take more than 4 hours.
	
	Recently, the rapid development of blockchain technology has enabled blockchain-based applications in various areas, including Internet-of-Things, healthcare, military, and service providers. In particular, thanks to its advantages of low latency and negligible computational requirement, the PoS consensus mechanism has emerged to be an effective solution to data management in networks consisting of devices with limited computational capacity~\cite{PoS}. Therefore, in this paper, we propose BlockRoam, a PoS blockchain solution to address the high delay problem in existing roaming systems.
	
	\subsection{Related Work}
\begin{figure}[ht]
		\includegraphics[width=.5\textwidth]{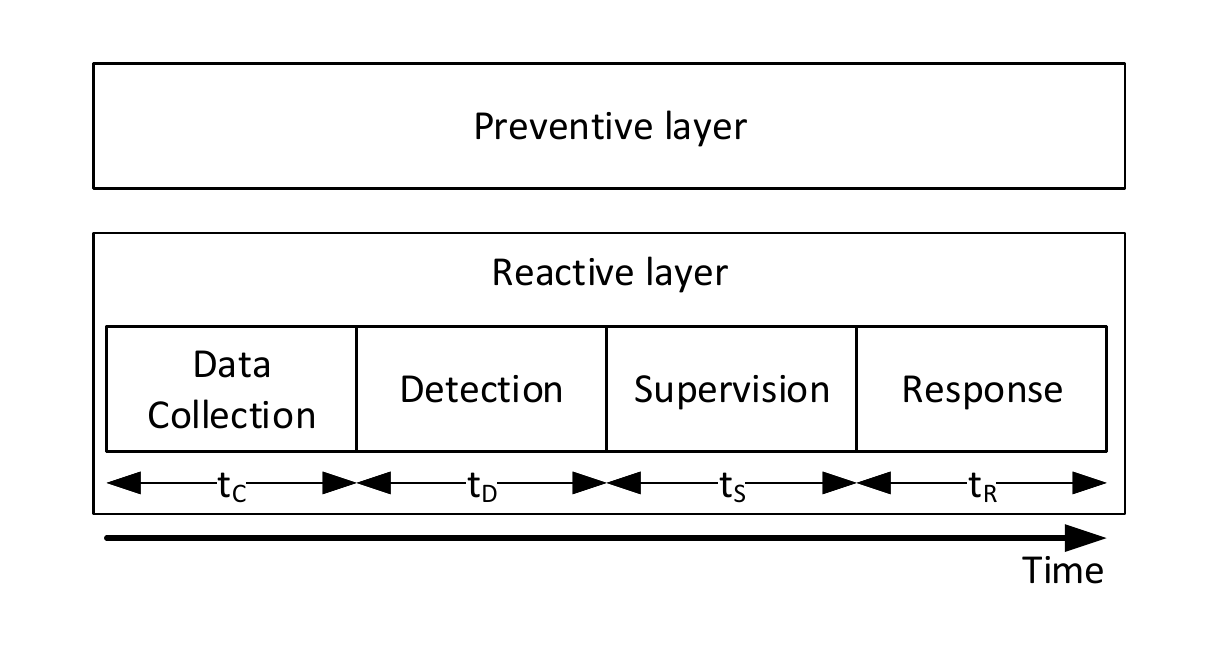}
		\centering
		\caption{Illustration of the current fraud protection system.}
		\label{Fig:currentfraud}
	\end{figure}

	Typically, a roaming fraud protection system consists of preventive and reactive layers as illustrated in Fig.~\ref{Fig:currentfraud}~\cite{fraud}. The preventive layer prevents fraud perpetration by validating subscribers' authentication, auditing subscribers' credit, limiting services duration, and so on. Although these measures can help to mitigate roaming frauds, they have a negative impact on the Quality-of-Service provided to the subscribers, e.g., frequent validation and service limitation will lower customer satisfaction. The reactive layer typically consists of four main stages to detect and react to roaming fraud attacks. The roaming data, e.g., service records, exchanged between MSPs is first collected at the data collection stage and processed at the fraud detection stage to detect potential fraud cases~\cite{fraud}. Each case is then supervised manually in the supervision stage. The service usage is terminated if a fraud attack is confirmed at the response stage. Among these stages, data collection is often the bottleneck in the roaming fraud protection system. Techniques employed at this stage can only support data collection in near real-time with a limited number of subscribers, e.g., Fraud Information Gathering System~\cite{FIGS}, or shorten the data exchanging delay to 4 hours, e.g., Near Real Time Roaming Data Exchange~\cite{ntrde}. Due to the sequential nature of the system, other stages cannot be activated if the data has not been collected. Consequently, although fraud attacks such as SIM cloning can also perpetrate locally in the HPMN, their consequences are much more severe in the roaming scenario due to the delay in data exchange, e.g., it takes up to 18 hours on average before an international roaming fraud attack can be stopped with the current system~\cite{loss2}.
	
	With outstanding performance in data integrity, decentralization, and privacy-preserving, blockchain has been emerging to be a secure and effective solution for data management in many decentralized networks. As a result, blockchain-based solutions for mobile roaming have been introduced recently by some organizations, e.g., IBM~\cite{IBM}, Deutsche Telekom and SK Telecom~\cite{DT}, and Enterprise Ethereum Alliance~\cite{EEA}, focusing on identity management, automating billing processes, and fraud prevention. In particular, these solutions focus on developing blockchain's asymmetric keys and digital signatures to manage subscriber identities and propose smart contracts to set up roaming pacts and automate billing processes. With enhanced identity management and automatic billing, fraud attacks can be significantly reduced. However, most of these solutions are still at the early stage of development and are facing several technical challenges. 
	
	Specifically, most of current blockchain-based data management systems often employ the Proof-of-Work (PoW) consensus mechanism, e.g., Bitcoin~\cite{Bitcoin}. However, the PoW mechanism consumes massive amounts of energy, e.g., the Bitcoin network's energy consumption is higher than that of many countries~\cite{energy}. Moreover, PoW-based networks often take a long time to reach consensus, e.g. one hour on average~\cite{PoS}. Thus, a new consensus mechanism, namely Proof-of-Stake (PoS), has been developed with significant advantages over the PoW mechanism, including reduced energy consumption and delay~\cite{PoS}. Recently, a PoS-based blockchain network, namely Bubbletone~\cite{bbt}, has been introduced for MSPs to address roaming fraud problems. Using the PoS-based consensus mechanism and smart contracts, the blockchain-based Bubbletone system provides a general platform for various MSP-to-MSP and MSP-to-subscriber interactions in the roaming environment. Nevertheless, the consensus mechanism design is not thoroughly discussed in~\cite{bbt}.
	
	In addition, more users (e.g., mobile subscribers) participate in a PoS-based blockchain network means better the performance and security of the network are. Thus, it is important to incentivize more users to participate in the network. In current PoS-based blockchain systems, some stakes, e.g., network tokens, are paid to the users as a reward for consensus participation. However, a user with a few stakes is less likely to receive the reward. Moreover, some blockchain networks such as~\cite{bbt} impose a high stake requirement for consensus participation. Consequently, the stakeholders, i.e., subscribers, are inclined to join a stake pool (formed by MSPs) to earn more rewards. Furthermore, a stake pool can earn profits from the investments of the stakeholders by charging a portion of each stakeholder's reward~\cite{PoS}. As a result, the formation of a stake pool can be beneficial if it can incentivize more subscribers and MSPs to join the network. Therefore, the design of stake pool and network parameters has a significant impact on the performance of a blockchain network, yet studies on this topic are still limited. The stake pool formation in PoS-based blockchain networks was analyzed in our previous work in~\cite{PoS}. However,~\cite{PoS} only considers the investment strategies of the users while the stake pool's pricing policy is assumed to be static. In practice, however, the pool has to design its pricing policy to maximize the profits while attracting more investments from the stakeholders.  
	
	\subsection{Contributions and Paper Organization}
	The main contributions of this paper are briefly summarized as follows:
	\begin{itemize}
		\item We propose BlockRoam, an effective blockchain-based roaming service management system to provide a transparent, secure, and automatic platform for data exchanging between the MSPs as shown in Fig.~\ref{Fig:fig2}. In particular, by employing the PoS consensus mechanism, BlockRoam can achieve a delay of fewer than 3 minutes as will be shown later in Section~\ref{sec:analysis}, which is much lower than the 4-hour delay of traditional roaming management systems. In addition to the reduced latency, BlockRoam can automate various roaming processes with the help of smart contracts~\cite{SC}, and thus roaming frauds can be significantly reduced. Moreover, the MSPs often rely on Data Clearing Houses (DCHs) to process and exchange data, which incurs additional costs~\cite{fraud}. In our proposed system, the transactions are stored in the blockchain and processed by smart contracts, and thus the service fees for DCHs can be eliminated. Furthermore, the privacy and security of the subscribers in BlockRoam are significantly enhanced thanks to the blockchain’s advanced cryptography techniques~\cite{wangsurvey}.
		\begin{figure*}[ht]
			\includegraphics[width=\textwidth]{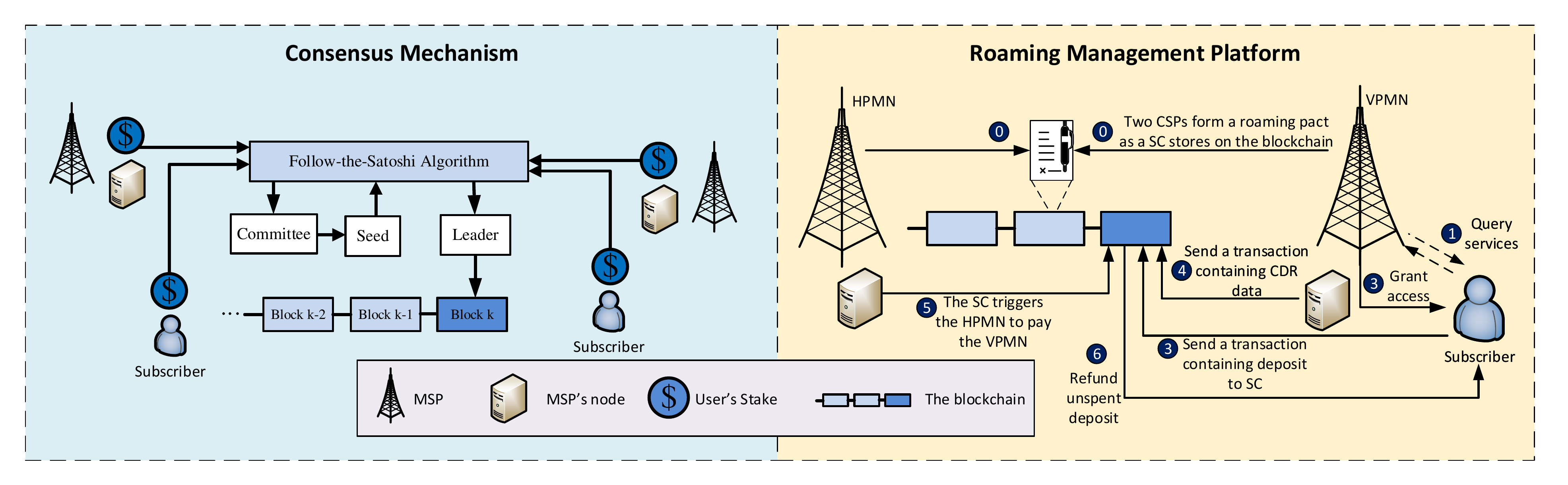}
			\centering
			\caption{Illustration of the proposed BlockRoam system.}
			\label{Fig:fig2}
		\end{figure*}
		\item We analyze the security performance of BlockRoam and prove that BlockRoam can meet strict security requirements of a blockchain-based system with improved reliability. We also show that BlockRoam can successfully prevent a wide variety of attacks, including double-spending, grinding, nothing-at-stakes, bribe, transaction denial, and long-range attacks. Moreover, we perform intensive performance analysis on real blockchain networks to show that the efficiency of proposed BlockRoam can be further improved by incentivizing more users to contribute to the network. This is also the main motivation for us to develop an economic model for the BlockRoam system.
		\item We introduce an economic model based on the Stackelberg game theory in order to jointly maximize the profits of the stake pool and the stakeholders. As a result, the stakeholders are incentivized to contribute more to BlockRoam. By analyzing utility functions of the stake pool and stakeholders, we develop a Mixed Integer Linear Programming model to find the Stackelberg equilibrium of our proposed game. We also propose an effective method that can guarantee to achieve the unique equilibrium for this game. The proposed economic approach can help the stake pool to obtain the optimal pricing policy and the stakeholders to find the best investment strategies. 
		\item Extensive simulation has been performed to evaluate the performance of our game theoretic model and the relations between the stake pool and stakeholders. We also examine the influence of important parameters on the outcome of the game. The results are especially crucial in designing appropriate parameters (e.g., total network stakes, pool fees, and rewards) for the stake pool to maximize its profit and attract more stakeholders to the network.
	\end{itemize}
	
	The remainder of this paper is organized as follows. We first provide the background about current mobile roaming systems and blockchain technology and introduce BlockRoam in Section~\ref{sec:system model}. We then analyze the security and performance of BlockRoam in Section~\ref{sec:analysis}. After that, we formulate and analyze the stake pool and stakeholders game in Section~\ref{sec:stakepool}. Finally, simulations and numerical results are presented in Section~\ref{sec:simu}, and conclusions are summarized in Section~\ref{sec:Sum}.

	%\newpage
	%=============================================================
	%=============================================================
	\section{Background and System Model}
	\label{sec:system model}

	%++++++++++++++++++++++++++++++++++++++++++++++
	%++++++++++++++++++++++++++++++++++++++++++++++
	\subsection{Current Roaming Systems}
\begin{figure}[ht]
		\includegraphics[width=.5\textwidth]{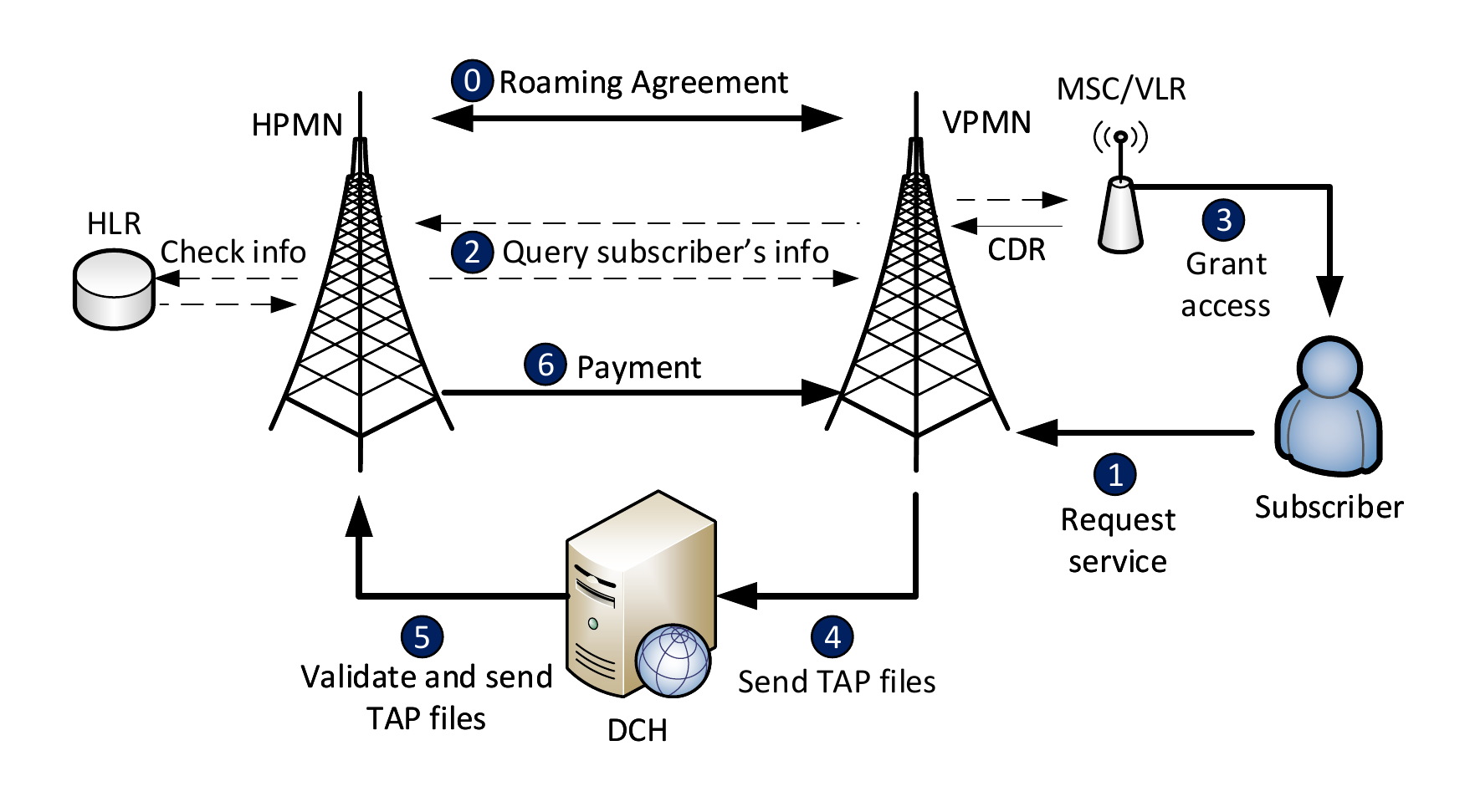}
		\centering
		\caption{Illustration of a typical roaming system~\cite{fraud}.}
		\label{Fig:fig1}
	\end{figure}

	The current roaming system is illustrated in Fig.~\ref{Fig:fig1}~\cite{fraud}. In the current system, firstly, a roaming pact is established between two MSPs. Then, when a subscriber wants to use services from its HPMN while being in the service area of the VPMN, the subscriber sends a request to the VPMN. Then, the VPMN queries the HPMN about the services that the subscriber has subscribed to. This information is stored in the Home Location Register (HLR) database of the HPMN. If the subscription information is correct, the VPMN will provide the subscriber access to the corresponding services (e.g., voice or data service) through the Mobile Switching Center/Visited Location Register (MSC/VLR). The Call Detail Records (CDRs) are then sent to both networks where the CDRs are processed for subscription billings and invoices generation. Afterward, the VPMN sends a Transfer Account Procedure (TAP) file which contains the CDR information to the HPMN. Usually, there is a Data Clearing House (DCH) company acting as a middleman, which validates and transmits the TAP files for the VPMN. Once the HPMN receives the TAP files, it will pay the VPMN in accordance with the roaming pact~\cite{fraud}.
	
	Fraud attacks in roaming occur when a subscriber gains access to the roaming services, but the HPMN is unable to charge the subscriber for the services provided. In this case, the HPMN still has to pay the VPNM for the facilities provided during the roaming process, which may result in significant financial loss. For example, a fraudulent SIM can use up to 18 hours of service on average, and in some incidents, the loss rate is up to \euro40,000 per hour~\cite{loss2}. The current roaming system is vulnerable to roaming fraud attacks mainly because of the delay in data exchanging between the HPMN and the VPMN. Even with the Near Real Time Roaming Data Exchange scheme~\cite{ntrde}, the data exchange can be delayed up to 4 hours, and thus it may take a long time to detect and determine the fraud. Even if the fraud is found, it is still difficult for the HPMN to response as it does not have direct control over the VPMN's facilities~\cite{fraud}.
	
	\subsection{Blockchain Fundamentals}
	A blockchain is a sequence (chain) of blocks, where each block consists of data (transactions) shared among users in the network. When a transaction is generated by a user, it will be first verified by miners, i.e., nodes who participate in the consensus process, to verify the transaction. After the transaction is verified and added to a new block, the block will be broadcast to the rest of the nodes in the network. Based on the distributed consensus mechanism, a block will be selected from all the blocks proposed by the miners to append to the chain~\cite{wangsurvey}. Besides the transactions, a block also contains a hash pointer created by the hash functions which map all the block contents and the last block's pointer to the current block's pointer. Therefore, any change in previous blocks will result in a different hash value in the next one, and it can be traced back to the first block of the chain. As a result, the whole blockchain is tamper-evident, i.e., any attempt to alter the previous blocks can be immediately detected. This is one of the most crucial advantages of blockchain technology compared to other security mechanisms. Another advantage is that a blockchain network is decentralized, and thus there is no single point of failure, i.e., the network's operation is ensured even when some nodes are failed. In contrast, for the current roaming system, if the DCH is failed, the CDRs and TAP files cannot be transmitted, and in this case, the whole system will stop working.
	
	A smart contract is a program stored in the blockchain network consisting of a set of rules created by users. If the rules are satisfied, the contract will automatically be enforced by the consensus mechanism. The content of a smart contract is visible to all network users, thus transparency is ensured~\cite{SC}. For example, an HPMN and a VPMN can negotiate with each other and make a smart contract on the blockchain, which is triggered when a transaction with CDR data is sent to the smart contract address. Then, when the transaction is verified and added into the blockchain, all consensus participants execute the contract code and trigger the events according to the terms of agreement written in the contract, e.g., the HPMN automatically pays the VPMN as per their agreement. 
	
	The distributed consensus mechanism is the backbone of a blockchain network, which governs most of the blockchain's operations and ensures that once the data is stored in a block, it is extremely difficult to be altered without the consensus of most of the nodes in the network. Currently, most of the blockchain networks have been employing the PoW consensus mechanisms. In the PoW, the users compete with each other in a solution searching procedure where a user with higher computational power may have higher opportunities to be the block winner who will add a new block to the chain and receive the reward. This competition leads to the waste of energy in PoW-based blockchain networks. Moreover, PoW-based blockchain networks often experience high delays in reaching consensus due to security reasons. This makes PoW consensus mechanisms inappropriate to implement in mobile roaming systems requiring low delay for fraud prevention. 
	
	Unlike the PoW, each block in PoS-based blockchain networks is dedicated to an authorized participant (leader) for mining in advance based on stakes of stakeholders in the network. This mechanism has many advantages over the PoW, including lower energy consumption and delay, and thus PoS-based blockchain applications can be employed effectively in networks with thousands of users~\cite{PoS}. Currently, there are some variations of PoS mechanisms. Some of them, such as Ouroboros~\cite{Ouroboros}, Casper~\cite{Casper}, and Tendermint~\cite{Tendermint}, employ a committee of several leaders instead of a single leader as in Chain-of-Activity~\cite{CoA} and Proof-of-Activity~\cite{PoA}. There are also several variations of the leader selection algorithm such as the Follow-the-Satoshi (FTS) algorithm~\cite{Ouroboros,Tendermint, Casper} and the cryptographic sortition algorithm~\cite{Algorand}. In some PoS networks, every new block that is added to the chain will be voted to confirm immediately, i.e., immediate finality~\cite{Tendermint,Algorand}, whereas in other networks, a block is confirmed after several new blocks are added to the chain, i.e., delayed finality~\cite{Ouroboros,Casper}. To penalize malicious behaviors, before a new block is created, some PoS networks~\cite{Casper,Tendermint,CoA} require the leader to make a deposit which is confiscated if the leader behaves maliciously. The main characteristics, including both advantages and limitations of the considered designs, are summarized in Table~\ref{tab:design}.
	
	\begin{table*}[]
		\centering
		\caption{PoS consensus mechanism designs comparison.}\label{tab:design}
		\begin{tabular}{c|l|l|l|}
			\cline{2-4}
			& \multicolumn{1}{c|}{\textbf{Characteristic}}                                             & \multicolumn{1}{c|}{\textbf{Advantages}}                  & \multicolumn{1}{c|}{\textbf{Limitations}}             \\ \hline
			\multicolumn{1}{|c|}{\multirow{2}{*}{\textbf{Committee~\cite{Ouroboros}}}}                & \multirow{2}{*}{Several committee members decide}               & \multirow{2}{*}{More secure than if}                             & \multirow{2}{*}{Longer voting time}                  \\[3pt]
			\multicolumn{1}{|c|}{}                                                  & seeds and confirm blocks                                                                                          &  there is no committee                                                          &               \\ \hline
			\multicolumn{1}{|c|}{\multirow{2}{*}{\textbf{No committee~\cite{PoA}}}}             & \multirow{2}{*}{One leader decides block}                                                & \multirow{2}{*}{Lower block time}                  & \multirow{2}{*}{Less secure than if}                         \\[3pt]
			\multicolumn{1}{|c|}{}                                                  &                                                                                          &                                                           & there is a committee                                                      \\ \hline \hline
			\multicolumn{1}{|c|}{\multirow{2}{*}{\textbf{FTS~\cite{Ouroboros}}}}                     & \multirow{2}{*}{Take seeds as input,}                                 & \multirow{2}{*}{None compared to } & \multirow{2}{*}{Leader is known }           \\[3pt]
			\multicolumn{1}{|c|}{}                                                  &output token index                                                                                          &cryptographic sortition                                                           &in advance                                                       \\ \hline
			\multicolumn{1}{|c|}{\multirow{2}{*}{\textbf{Cryptographic sortition~\cite{Algorand}}}} & \multirow{2}{*}{Take seeds and private key as input, }       & \multirow{2}{*}{Leader cannot be }        & \multirow{2}{*}{No advantage if } \\[3pt]
			\multicolumn{1}{|c|}{}                                                  & and output token index and proof                                                                                          &known in advance                                                           &there is a committee                                                       \\ \hline \hline
			\multicolumn{1}{|c|}{\multirow{2}{*}{\textbf{Delayed finality~\cite{Ouroboros}}}}        & \multirow{2}{*}{Block is confirmed after several }                                 & \multirow{2}{*}{Higher block time}                        & \multirow{2}{*}{Higher }             \\[3pt]
			\multicolumn{1}{|c|}{}                                                  &  blocks deep in the chain                                                                                         &                                                           &confirmation time                                                       \\ \hline
			\multicolumn{1}{|c|}{\multirow{2}{*}{\textbf{Immediate finality~\cite{Tendermint}}}}      & \multirow{2}{*}{Block is confirmed immediately }                                          & \multirow{2}{*}{Lower confirmation time}                  & \multirow{2}{*}{Lower block time}                     \\[3pt]
			\multicolumn{1}{|c|}{}                                                  &by voting                                                                                          &                                                           &                                                       \\ \hline \hline
			\multicolumn{1}{|c|}{\multirow{2}{*}{\textbf{Reward mechanism~\cite{Ouroboros}}}}        & \multirow{2}{*}{Block reward for leader}                                                 & \multirow{2}{*}{Incentivize }      & \multirow{2}{*}{None}                                 \\[3pt]
			\multicolumn{1}{|c|}{}                                                  &                                                                                          &consensus participation                                                           &                                                       \\ \hline
			\multicolumn{1}{|c|}{\multirow{2}{*}{\textbf{Penalty mechanism~\cite{Casper}}}}       & \multirow{2}{*}{Leader has to make deposit which is } & \multirow{2}{*}{Mitigate several }        & \multirow{2}{*}{None}                                 \\[3pt]
			\multicolumn{1}{|c|}{}                                                  &confiscated for malicious behaviors                                                                                          &types of attacks                                                           &                                                       \\ \hline
		\end{tabular}
	\end{table*}
	In our proposed blockchain network, we employ a dynamic committee selected based on the stake distribution, as it is more secure than the single leader case. In committee-based blockchain networks, the committee members are responsible for the consensus process for a certain period of time, and thus they are known in advance. Consequently, there is no advantage in using the cryptographic sortition over the FTS algorithm. Therefore, we choose the FTS for the leader selection process. In our proposed system, the committee members do not vote to confirm each block. This is because the roaming environment may involve thousands of users, and thus a low block time, i.e., the time it takes to add a new block to the chain, is more desirable than a low confirmation time. To incentivize participation in the network, a leader can receive the block reward, e.g., a fixed number of tokens, each time the leader adds a block to the chain. The leader is also required to make a deposit which will be confiscated for malicious behaviors. 
	\subsection{BlockRoam}
	\subsubsection{Network Model} Our proposed blockchain-based system consists of two main components, namely the roaming management platform and the consensus mechanism as illustrated in Fig.~\ref{Fig:fig2}. The roaming management platform supports complex interactions between the users, automates various roaming processes, and provides a universal currency, i.e., blockchain network tokens, for payments. In addition to the roaming processes, the network can also take part in the consensus mechanism to maintain the network's operations and security, store data (e.g., roaming pacts, subscriber information, and transaction history), and execute roaming processes such as payments and processing CDRs.  
	
	\subsubsection{Roaming Management Procedure} The roaming process, the main procedure of the roaming management platform, consists of seven main steps as follows:
	\begin{itemize}
		\item \textit{Step 0:} Two MSPs form a roaming pact consisting of tariff plans for services offered to the subscribers and the payment agreement between two MSPs. This roaming pact is made in the form of a smart contract and stored in the blockchain.
		\item \textit{Step 1:} When a subscriber (roamer) wants to use services from its HPMN, the subscriber queries the VPMN and receives available tariff plans as per the roaming agreement between the VPMN and the HPMN.
		\item \textit{Step 2:} If the subscriber agrees to use the service, the subscriber sends a transaction containing a sufficient amount of money (in form of digital tokens) to the smart contract's address.
		\item \textit{Step 3:} When the transaction is verified and sent successfully, the VPMN will grant the subscriber access to roaming facilities.
		\item \textit{Step 4:} When the subscriber finishes its roaming service, the VPMN sends a transaction to the smart contract's address, which consists of the CDR data of the provided service.
		\item \textit{Step 5:} The smart contract then automatically calculates the subscriber's service fee and sends it to the HPMN. The smart contract also triggers a transaction from the HPMN to the VPMN for payment of the service.
		\item \textit{Step 6:} Finally, the smart contract sends the unused tokens to the subscriber.
	\end{itemize}  
	
	\subsubsection{Consensus Mechanism}
	In our proposed blockchain network, the stake of a user corresponds to the number of network tokens the user currently holds. The consensus mechanism employs a dynamic committee selected based on the stake distribution. Time is divided into epochs during each of which, the committee members participate in a 3-phase coin-tossing protocol to create seeds for the FTS algorithm~\cite{Ouroboros}. Using the seeds created by the committee, the FTS algorithm selects the leaders and committee members for the next epoch. The probability $P_i$ that user $i$ is selected by the FTS algorithm in a network of $N$ users is
	\begin{equation}
	\label{eq:PoS}
	P_i=\dfrac{s_i}{\sum_{n=1}^{N}s_n},
	\end{equation}
	where $s_i$ is the number of stakes of user $i$. This means that the more stakes a user holds, the higher chance it can be selected to be the leader. Each epoch is further divided into slots, and in each slot, a designated leader adds a new block to the chain. To incentivize participation in the network, a leader will receive a fixed number of tokens, when the leader adds a new block to the chain. The leader is also required to make a deposit that will be locked during its designated epoch to prevent nothing-at-stake, bribe~\cite{PoS}, and transaction denial attacks~\cite{Ouroboros}. The stakes of committee members are also locked during the epoch to prevent long-range attacks~\cite{Casper}.   
	\subsubsection{Benefits}
	BlockRoam has the following advantages over the traditional roaming system:
	\begin{itemize}
		\item \textit{Roaming fraud prevention:} The main obstacle to prevent and react to fraud attacks is the significant delay in data exchange, i.e., up to 4 hours. Our proposed system employs the PoS mechanism to speed up the data exchanging process, e.g., approximately 3 minutes on average as later shown in Section~\ref{sec:analysis}, and thus fraud attacks can be detected much earlier. Moreover, by using smart contracts, the billing process is executed right after the service usage finished. As a result, roaming fraud can be significantly mitigated.
		\item \textit{Cost saving:} In our proposed system, the CDRs are stored in the blockchain and processed by smart contracts. Therefore, the DCHs are no longer needed, and thus the middleman fees are eliminated. Moreover, our system automates various processes, such as subscribers billing and HPMN payments, which can further reduce operational costs. Furthermore, our system's energy consumption is negligible compared to that of PoW-based systems, and thus our energy cost is much lower.
		\item \textit{Security and privacy:} Using cryptographically secure mechanisms, the privacy and security of the subscribers can be significantly improved. Each subscriber in the network uses a pair of public and private keys for identification and verification. The network only needs the subscriber's digital signature which can be easily verified and almost impossible to forge. This also protects the anonymity of the subscribers, as the subscriber's real-life identity is completely unrelated to the network identity.
	\end{itemize}
	\section{Security and Performance Analysis}
	\label{sec:analysis}
	\subsection{Security Analysis}
	\subsubsection{Blockchain Properties}
	To maintain the blockchain's operations and security, a consensus mechanism must satisfy the following properties~\cite{sec}: \begin{itemize}
		\item \textbf{Persistence:} Once a transaction is confirmed by an honest user, all other honest users will also confirm that transaction, and its position is the same for all honest users. 
		\item \textbf{Liveness:} After a sufficient period of time, a valid transaction will be confirmed by all honest users.
	\end{itemize}
	In our proposed system, persistence ensures that once a transaction is confirmed, it cannot be reverted. Without persistence, a fraudster can use the roaming services for free. For example, a fraudster can perform a double-spending attack by firstly sending a transaction $Tx_1$ to the smart contract. Then, after the VPMN has granted the fraudster access to the roaming service, the fraudster broadcasts a transaction $Tx_2$ which sends the tokens of $Tx_1$ to another address (e.g., the fraudster's second account). If $Tx_1$ has not been confirmed, $Tx_2$ is still valid and may be confirmed by honest users.
	
	While the persistence property ensures data immutability, the liveness property ensures that every valid transaction will eventually be included in the chain. Without liveness, an attacker might successfully block every transaction coming from the MSP, and consequently, the roaming process cannot commence. It has been proven in~\cite{sec} that the persistence and liveness properties are ensured if the consensus mechanism satisfies the following properties:
	\begin{itemize}
		\item \textbf{Common prefix (CP) with parameter $\kappa \in \mathbb{N}$:} For any pair of honest users, their versions of the chain $\mathcal{C}_1,\mathcal{C}_2$ must share a common prefix. Specifically, assuming that $\mathcal{C}_2$ is longer than $\mathcal{C}_1$, removing $\kappa$ last blocks of $\mathcal{C}_1$ results in the prefix of $\mathcal{C}_2$. 
		\item \textbf{Chain growth (CG) with parameter $\varsigma \in \mathbb{N}$ and $\tau\in (0,1]$:} A chain possessed by an honest user at time $t+\varsigma$ will be at least $\varsigma\tau$ blocks longer than the chain it possesses at time $t$.
		\item \textbf{Chain quality (CQ) with parameter $l \in \mathbb{N}$ and $\mu \in (0,1]$:} Consider any part of the chain that has at least $l$ blocks, the ratio of blocks created by the adversary is at most $1-\mu$.
	\end{itemize}
	Since our consensus mechanism shares many similarities with the Ouroboros consensus mechanism~\cite{Ouroboros}, we have the following important parameters: 
	\begin{itemize}
		\item Common prefix violation probability $\rm Pr_{CP}$: Let $\gamma$ be the proportion of the total network stakes controlled by honest users. The probability that our consensus mechanism violates the common prefix property with parameter $\kappa \in \mathbb{N}$ over an epoch of $\rho$ slots is no more than ${\rm e}^{-\Omega(\sqrt{\kappa})+\ln \rho}$, where the constant hidden by $\Omega(.)$ depends only on $\gamma$. 
		\item Chain growth (CG) violation probability $\rm Pr_{CG}$: The probability that our consensus mechanism violates the chain growth property with parameters $\tau, \varsigma$ over an epoch of $\rho$ slots is no more than ${\rm e}^{-\Omega(\epsilon^2\varsigma)+\ln \rho}$.
		\item Chain quality (CQ) violation probability $\rm Pr_{CQ}$: The probability that our consensus mechanism violates the chain quality property with parameters $l, \mu$ over an epoch of $\rho$ slots is no more than ${\rm e}^{-\Omega(\epsilon^2\gamma l)+\ln \rho}$.
	\end{itemize}
	In~\cite{Ouroboros}, the bound of $\rm Pr_{CP}$ is proven based on the probability that the adversary can create a fork (i.e., a different version of the chain) longer than the honest one. However, this is a conservative approach, because even if the adversary can create a longer fork, it does not necessarily mean that the entire honest fork is abandoned. In particular, we will prove that our consensus mechanism can achieve a new bound of $\rm Pr_{CP}$ based on the following properties:
	\begin{itemize}
		\item \textbf{$A_1$:} An honest user will create exactly one block for each slot that the user is the leader. 
		\item \textbf{$A_2$:} The list of leaders is known by every honest user at any time.
		\item \textbf{$A_3$:} An honest user, when received different forks, will adopt the longest valid fork, i.e., the longest fork that has no conflicting blocks and each block is signed by a designated leader. 
	\end{itemize}
	Then, we prove a new bound of $\rm Pr_{CP}$ in the following theorem.
	\begin{theorem}
		Suppose properties $A_1, A_2$ and $A_3$ are satisfied, the probability that BlockRoam's consensus mechanism violates the common prefix property with parameter $\kappa \in \mathbb{N}$ is less than or equal to $(1-\gamma)^\kappa$. 
	\end{theorem}
	\begin{IEEEproof}
		Suppose properties $A_1$, $A_2$, and $A_3$ are satisfied, any fork created by the adversary must include all the blocks created by the honest users. This is because if an honest user does not change its block, then the adversary can either adopting the block in the fork or replace it by another block. However, as the list of leaders is known, the adversary must include the honest block in the fork. Otherwise, it will create an invalid fork that will be rejected based on property $A_3$. Moreover, any change in a block's content results in a different block's hash, and the block's hash is linked to its previous block. Thus, the part of the chain from the first block to the latest honest block is confirmed by every honest user. As a result, the adversary can only create forks with $\kappa$ last blocks different from the honest fork if it is elected leader for $\kappa$ consecutive blocks. Since $(1-\gamma)$ is the ratio of adversarial stakes in the total network stakes, the probability that the adversary is elected leader for $\kappa$ consecutive blocks is 
		\begin{equation}
		\label{prob}
		Pr_{CP}=(1-\gamma)^\kappa,
		\end{equation}
		which is also the probability that the common prefix property is violated.
	\end{IEEEproof}
	Properties $A_1$ and $A_3$ can be easily satisfied if all the honest users follow the consensus mechanism. Property $A_2$ can be ensured by conducting the coin-tossing protocol at the beginning of each epoch, instructing the honest users to broadcast the leader list of the next epoch during the current epoch, and requiring an honest user to be online at least once each epoch (this requirement is reasonable since an epoch of~\cite{Ouroboros} lasts for 5 days).
	\subsubsection{Roaming Fraud Protection Ability}
	To evaluate the roaming fraud protection ability of our system, we focus on the average resolution time $t_{total}$, i.e., the average time between the occurrence of a roaming fraud attack and the execution of the responses to the attack. As observed in Fig.~\ref{Fig:currentfraud}, $t_{total}$ is the sum of every stage's duration at the reactive layer, i.e., $t_{total}=t_C+t_D+t_S+t_R$. Since our proposed system can achieve a much lower $t_C$ compared to the traditional roaming system, i.e., approximately 3 minutes (as later shown in Section~\ref{perform}) compared to 4 hours, the $t_{total}$ of our system is nearly 4 hours shorter than that of the traditional roaming system.
	\subsubsection{Blockchain Attacks Mitigation}
	\label{attack}
	In the following Theorem, we prove that our proposed BlockRoam can also be able to mitigate and prevent a variety of emerging blockchain attacks such as double spending, grinding, bribe, nothing-at-stakes, and long-range attacks. 
	\begin{theorem}
		BlockRoam can mitigate double-spending, grinding, nothing-at-stakes, bribe, transaction denial, and long-range attacks as long as the adversary does not control more than 50\% total network stakes.
	\end{theorem}
	\begin{IEEEproof}
		In a double-spending attack, the adversary attempts to revert a transaction by adding a conflicting transaction to the blockchain after the original transaction is confirmed. It is straightforward to see that this attack cannot be successful if the common prefix property is not violated. 
		
		In grinding attacks and nothing-at-stake attacks, the adversary creates multiple blocks to influence the seeds of the leader selection process or revert some blocks in the chain. More specifically, grinding attacks target the blockchain where the seeds for leader selection are the previous blocks' headers. However, the seeds for leader selection are created by the committee in BlockRoam, and thus grinding attacks are mitigated. Moreover, although the adversary can create forks, nothing-at-stakes attacks do not affect the network's security as long as the common prefix property is not violated. Furthermore, the adversary's deposit will be confiscated if the adversary signs different blocks for the same time slot.
		
		In bribe attacks, the adversary can bribe the leaders to create specific blocks, e.g., to support other types of attacks such as double-spending or transaction denial. In the context of roaming, bribe attacks may cause severe financial loss. For example, an adversary can perform a bribe attack to support a transaction denial attack, i.e., bribe the leaders to not include any transaction made from a certain MSP, and consequently that MSP cannot process any roaming request. In this case, the deposits will be confiscated, which significantly increases the costs of bribe attacks and transaction denial attacks. 
		
		In a long-range attack, a committee member immediately sells its stakes at the beginning of its designated epoch, and thus it can behave maliciously for the rest of the epoch without consequences. Our system can mitigate this attack by locking committee members' stakes during their designated epoch. 
	\end{IEEEproof}
	When the adversary controls more than 50\% of the total network stakes, both the persistence and liveness properties are no longer guaranteed~\cite{Ouroboros}. Consequently, attacks such as double-spending, nothing-at-stakes, and transaction denial attacks can no longer be mitigated.
	\subsection{Performance Analysis}
	\label{perform}
	In Table~\ref{tab:Compare}, we examine and compare the transaction confirmation times under different adversarial ratio (percentage of stakes in PoS or computational power in PoW that the adversary controls) of a PoW blockchain network (Bitcoin), a PoS network with delayed finality (Cardano), and BlockRoam. The transaction confirmation time is the time it takes to reach a common prefix violation probability $Pr_{CP} \leq 0.1 \%$. Based on (\ref{prob}), $\kappa$ can be determined, and then $\kappa$ is multiplied with the slot time to calculate the transaction confirmation time. Our slot time is set to be 20 seconds (the same as that of Cardano~\cite{Cardanotime}). The transaction confirmation times of Bitcoin and Cardano are presented in~\cite{Ouroboros}.
	\begin{table}[]
		\small
		\centering
		\caption{Transaction confirmation times in minutes} 
		\begin{tabular}{|c|c|c|c|}
			\hline
			Adversarial ratio& Bitcoin                    &  Cardano       & BlockRoam           \\ 
			\hline
			0.10             & 50   & 5  &       1           \\ 
			\hline
			0.15             & 80             & 8 & 1.3\\ 
			\hline
			0.20           &     110           & 12         &  1.6                \\ 
			\hline
			0.25         & 150 &  18      &  1.6 \\ 
			\hline
			0.30 &  240                    &  31      & 2  \\
			\hline
			0.35 & 410 & 60  & 2.3  \\
			\hline
			0.40 & 890 & 148  & 2.6  \\
			\hline
			0.45 & 3400 & 663  & 3  \\
			\hline
		\end{tabular}
		\label{tab:Compare}
	\end{table}
	
	As observed in Table~\ref{tab:Compare}, the more stakes the adversary controls, the longer the transaction confirmation time is. Moreover, 51\% attack~\cite{Ouroboros} can break most of the PoW-based and PoS-based blockchain networks. Specifically, an adversary controlling more than 51\% of total computational power in a PoW-based network or 51\% of total stakes in a PoS-based network can successfully perform many attacks, including double-spending, nothing-at-stakes, and transaction denial attacks. Therefore, it is critical to attract more participants to our PoS-based blockchain system in order to increase the network's total stakes and prevent the adversary from controlling more than 50\% of network stakes. In the next section, we will introduce an effective economic model that can jointly maximize profits for the participants, encouraging them to participate in the network and thus improving the network's performance and security.
	
	\section{Economic Model}
	\label{sec:stakepool}
	\subsection{Stake Pools and Stakeholders}
	In a PoS-based blockchain network, the probability that an individual user (stakeholder) with a small number of stakes is selected to be the leader is low as shown in (\ref{eq:PoS}). Moreover, to participate in the consensus process, a stakeholder must always be connected to the network, which incurs an operational cost, e.g., \$40 to \$300 per month~\cite{cost}. Therefore, small stakeholders often pool their stakes together to increase their opportunities to be leaders and share operational costs, which results in the formation of stake pools, e.g.,~\cite{Pool1,Pool2,Pool3}. In BlockRoam, the stakeholders, e.g., the subscribers, might be more inclined to join the stake pool (e.g., formed by MSPs) to reduce their operational costs and have more stable incomes. A stake pool often charges a part of the stakeholder's profits for joining the pool, e.g., the Stakecube pool charges 3\% of each reward a stakeholder receives~\cite{Pool2}. In this section, we introduce an economic model using Stackelberg game in order to jointly maximize the profits of the stake pool and stakeholders, which is beneficial for MSPs and BlockRoam's operation and security.
	
	We consider a PoS-based blockchain network with one stake pool and $N$ stakeholders. The stakeholders have stake budgets $\mathbf{B} = (B_1, \ldots, B_N)$ and individual operational costs $\mathbf{C} = (C_1, \ldots, C_N)$. The stake pool has its own stake $\sigma$, and the pool defines a cost $c$ and a fee $\alpha$ in advance for users who are interested in participating in the pool. The pool's cost is charged for joining the pool and maintaining its operations. The pool's fee is the profit margin of the pool's owner, which usually ranges from 1\% to 9\%  in real-world stake pools, e.g.,~\cite{Pool1,Pool2,Pool3}. The stakeholders can use their budgets to invest $p_i$ stakes to the pool and $m_i$ stakes for self-mining (individually participate in the consensus process), such that $p_i+m_i\leq B_i$. Let denote $\mathcal{N}_p$ to be the set of stakeholders who invest in the pool, the probability $P^w$ that the pool is selected to be the leader and obtains a block reward $ R $ is proportional to the pool's stakes in the total network stakes, i.e.,
	\begin{equation}
	\label{eq:probpool}
	P^w=\dfrac{\sigma+\sum_{n \in \mathcal{N}_p}p_n}{\sigma+\sum_{n \in \mathcal{N}_p}p_n+\sum_{j=1}^{N}m_j}.
	\end{equation}
	After receiving the reward $R$, the pool calculates each stakeholder's reward $r^p_i$ based on the proportion $P^p_i$ of stakeholder $i$'s stakes in the total stakes of the pool, which is
	\begin{equation}
	\label{eq:probplayer}
	P^p_i=\dfrac{p_i}{\sigma+\sum_{n \in \mathcal{N}_p}p_n}.
	\end{equation}
	The pool then charges a fee for $\alpha$ percentage from each stakeholder's reward and a cost of $ c{\rm e}^{-p_i} $ before the reward is finally sent to each stakeholder. Since the cost decreases exponentially as the stakes increase, it encourages the stakeholders to invest more stakes to the pool. Thus, when a stakeholder $i$ invests $p_i$ stakes to the pool, the stakeholder's expected reward $r^p_i$ is given by
	\begin{equation}
	\label{eq:reward}
	\begin{split}
	r^{p}_i&=P^w P^p_i(1-\alpha)R-c{\rm e} ^{-p_i},\\
	&=\dfrac{p_i}{\sigma+\sum_{n \in \mathcal{N}_p}p_n+\sum_{j=1}^{N}m_j}(1-\alpha)R-c{\rm e}^{-p_i}.
	\end{split}
	\end{equation}
	
	In the case if the stakeholder $i$ uses $m_i$ stakes to self-mine, its expected reward is
	\begin{equation}
	\label{eq:reward2}
	r^{m}_i =\bigg(\dfrac{m_i}{\sigma+\sum_{n \in \mathcal{N}_p}p_n+\sum_{j=1}^{N}m_j}\bigg)R-C_i,
	\end{equation}
	where $\dfrac{m_i}{\sigma+\sum_{n \in \mathcal{N}_p}p_n+\sum_{j=1}^{N}m_j}$ represents the proportion of stakeholder $i$'s stakes in the total network stakes. Then, the profit of the pool can be calculated as follows:
	\begin{equation}
	\label{eq:pooltotal}
	\begin{split}
	U_p&=\dfrac{\sigma}{\sigma+\sum_{n \in \mathcal{N}_p}p_n+\sum_{j=1}^{N}m_j}R\\&+\sum_{i \in \mathcal{N}_p}\bigg( \dfrac{p_i\alpha}{\sigma+\sum_{n \in \mathcal{N}_p}p_n+\sum_{j=1}^{N}m_j}R+c{\rm e}^{-p_i}\bigg).    
	\end{split}
	\end{equation}
	The total profit of the pool consists of the profits from its own stakes, i.e., the first term in (\ref{eq:pooltotal}), and the costs and fees it charges the stakeholders, i.e., the second term in (\ref{eq:pooltotal}).
	\subsection{Stackelberg Game Formulation}
	In practice, a pool usually announces its cost and fee first, e.g., the fee to join the Stakecube pool can be found on its website~\cite{Pool2}. Based on that information, the stakeholders will decide how much to invest. As a result, the interaction between the stake pool and stakeholders can be formulated to be a single-leader-multiple-followers Stackelberg game~\cite{Game}. In this game, the leader is the stake pool who first announces its strategy, i.e., costs and fees to join the pool, and then the stakeholders, i.e., followers, will make their decisions, e.g., to invest to the pool or not. %We denote $\Psi_{-j}$ to be the set of all possible strategies of all players (followers and leader) except player $j$. Furthermore, we denote $\boldsymbol{\psi}_{-j} \in \Psi_{-j}$ to be a strategy set of all players except player $j$ in $\Psi_{-j}$. Then, a strategy ${\psi}'_{i}$ of player $j$ is defined to be strictly dominated by another strategy ${\psi}_{j}$ if:
	%\begin{equation}
	%\begin{aligned}
	%\label{eq:dominate}
	%U_j({\psi}'_{j},\boldsymbol{\psi}_{-j})<U_j({\psi}_{i},\boldsymbol{\psi}_{-j}), \forall \boldsymbol{\psi}_{-j} \in \Psi_{-j}.
	%\end{aligned}
	%\end{equation}
	%In other words, ${\psi}'_{j}$ is strictly dominated by ${\psi}_{j}$ if ${\psi}_{j}$ yields a better payoff $U_j$ given any possible strategies of the other players. In this case, the dominated strategies can be eliminated because follower $j$ has no reason to choose a strategy that always gives worse payoff. If there exists a ${\psi}_{j}$ which dominates all other possible strategies of follower $j$, ${\psi}_{j}$ is the dominant strategy. In the case where every player has a dominant strategy, the system can reach a dominant-strategy equilibrium because all the players will rationally choose their dominant strategies. Nevertheless, dominant-strategy equilibrium does not often exist in many games~\cite{Game}.
	
	We denote $s_p$ and $s_i$ to be the strategies of the leader and follower $i$, respectively. Furthermore, we denote $\mathcal{S}_i$ to be the set of all possible strategies of follower $i$. Then, the best response $s^*_i$ of a follower $i$ can be defined to be the strategy set which gives the follower the best payoff given a fixed strategy $s_p=(\alpha,c)$ of the leader, i.e.,
	\begin{equation}
	\label{eq:bestresponse}
	U_i(s^*_i,s_p)\geq U_i(s'_i,s_p), \forall s'_i \in \mathcal{S}_i.
	\end{equation}
	Based on the follower's best response, the Stackelberg strategy for the leader is a strategy  $s^*_p$ such that
	\begin{equation}
	\label{eq:leader}
	s^*_p= \argmax_{s_p} U_p(s_p,s^*_i).
	\end{equation}
	Then, the Stackelberg solution can be defined as the tuple $(s^*_p,s^*_i)$, and its corresponding utility tuple $(U^*_p, U^*_i)$ is the Stackelberg equilibrium of the game. To find the Stackelberg equilibrium, the game can be divided into two stages. At the first stage, the leader announces its strategy. Then, at the second stage, the followers determine their strategies based on the leader's strategy. In the following, the backward-induction-based analysis is carried out to examine the Stackelberg equilibrium of this game.
	
	\subsubsection{Follower strategy}In this game, a follower's possible strategies can be divided into four cases:
	\begin{itemize}
		\item \textit{Case 1:} Only invest stakes to the pool.
		\item \textit{Case 2:} Only invest stakes for self-mining. 
		\item \textit{Case 3:} Simultaneously invest stakes to the pool and for self-mining.
		\item \textit{Case 4:} Do not invest stakes to the PoS-based blockchain network. 
	\end{itemize}
	In Case 1 and 2, although the follower can invest using any number of stakes within its budget, we prove in Lemma 1 that a rational follower will always invest all its budget.
	\begin{lemma}
		Let $s'_i$ denote a strategy where follower $i$ invests less than its total budget, i.e., $m'_i+p'_i<B_i$, with corresponding utility $U'_i$, and $s_i$ is a strategy where follower $i$ invests all its budget, i.e., $m_i+p_i=B_i$, with corresponding utility $U_i$. For Case 1 and Case 2, we always have $U'_i < U_i$. 
	\end{lemma}
	\begin{IEEEproof}
		We consider Cases 1 and Case 2 separately as follows:
		\begin{itemize}
			\item \textit{Case 1:} When the follower only invests $p'_i<B_i$ stakes to the pool, its expected payoff $U'^1_i$ is equal to $r^p_i$ in (\ref{eq:reward}). Now, if the follower invests $p_i=B_i$ to the pool, its payoff can be determined as follows: 
			\begin{equation}
			\label{eq:P1}
			U^1_i=\dfrac{B_i(1-\alpha)}{\sigma+\sum_{n \in \mathcal{N}_p}p_n+\sum_{j=1}^{N}m_j}R-c{\rm e}^{-B_i}.
			\end{equation}
			Then, the difference in payoff between the two strategies is
			\begin{equation}
			\label{eq:D1}
			\begin{split}
			U^1_i-U'^1_i&=\dfrac{(B_i-p'_i)(1-\alpha)}{\sigma+\sum_{n \in \mathcal{N}_p}p_n+\sum_{j=1}^{N}m_j}R\\&+(c{\rm e}^{-p'_i}-c{\rm e}^{-B_i}),
			\end{split}
			\end{equation}
			which is always positive since $p'_i<B_i$.
			\item \textit{Case 2:} When follower $i$ only uses $m'_i<B_i$ stakes for self-mining, its payoff $U'^2_i$ is equal to $r^m_i$ in (\ref{eq:reward2}). If the follower self-mines with all its budget, the payoff is
			\begin{equation}
			\label{eq:payoff2}
			U^2_i=\dfrac{B_i}{\sigma+\sum_{n \in \mathcal{N}_p}p_n+\sum_{j=1}^{N}m_j}R-C_i.
			\end{equation}
			The different in payoff is then determined by:
			\begin{equation}
			\label{eq:D2}
			U^2_i-U'^2_i=\dfrac{B_i-m'_i}{\sigma+\sum_{n \in \mathcal{N}_p}p_n+\sum_{j=1}^{N}m_j}R,
			\end{equation}
			which is always positive since $m'_i<B_i$.
		\end{itemize}   
	\end{IEEEproof}
	
	Moreover, we prove in the following Lemma that, given the same stakes to invest, Case 3 always gives a worse payoff than Case 2, and thus a rational follower will never choose Case 3.
	\begin{lemma}
		Let $U^2_i$, $U^3_i$ denote the payoff of Case 2 and Case 3, respectively. If follower $i$ invests the same $\beta$ stakes in these two cases, i.e., $m^2_i=\beta$ and $m^3_i+p^3_i=\beta$, then Case 2 always gives a better payoff than Case 3, i.e., $U^2_i > U^3_i, \forall \alpha, c$. 
	\end{lemma}
	\begin{IEEEproof}
		The difference in payoff between Case 2 and 3 can be calculated by
		\begin{equation}
		\label{eq:diff1}
		\begin{split}
		U^2_i-U^3_i=&\dfrac{\beta}{\sigma+\sum_{n \in \mathcal{N}_p}p_n+\sum_{j=1}^{N}m_j}R-C_i\\&-\bigg(\dfrac{\beta-p^3_i+p^3_i(1-\alpha)}{\sigma+\sum_{n \in \mathcal{N}_p}p_n+\sum_{j=1}^{N}m_j}R\\&-C_i-c{\rm e}^{-p^3_i}\bigg),\\
		=&\dfrac{p^3_i\alpha}{\sigma+\sum_{n \in \mathcal{N}_p}p_n+\sum_{j=1}^{N}m_j}R+c{\rm e}^{-p^3_i},
		\end{split}
		\end{equation}
		which is always positive. 
	\end{IEEEproof}
	As a result, Case 3 can be removed from the strategy space of every follower. 
	
	In Case 4, the follower receives payoff $U^4_i=0$. Therefore, if follower $i$ has budget $B_i$ such that $r^p_i>0$ or $r^m_i>0$, the follower will invest stakes to the pool or to self-mining, i.e., switch to Case 1 and 2. If follower $i$ has $B_i$ such that $r^p_i<0$ and $r^m_i<0$, the follower will not participate in the consensus process, and thus it does not have any impact on the game. Since the network benefits from user participation, network parameters such as $R$ should be designed to encourage stakeholders with small budgets to participate.
	
	Since Case 3 and Case 4 are eliminated and the strategies investing less than the budget always give less payoffs in Case 1 and Case 2, the total network stakes becomes a constant, i.e., 
	\begin{equation}
	\sigma+\sum_{n \in \mathcal{N}_p}p_n+\sum_{j=1}^{N}m_j=\sigma+\sum_{i=1}^{N}B_i.
	\end{equation} 
	Then, the best response of a stakeholder (i.e., follower) can be determined by Theorem 3.
	\begin{theorem}
		A stakeholder's best response is to invest all stakes either to invest to the pool or for self-mining. 
	\end{theorem}
	\begin{IEEEproof}
		Since Case 4 does not have any impact on the game, it follows from Lemma 1 and Lemma 2 that a rational stakeholder will use all its budget either to invest to the pool or for self-mining. 
	\end{IEEEproof}
	
	Since $p^*_i=B_i-m^*_i$, the best response can be deduced from either $p^*_i$ or $m^*_i$. Therefore, from now on, we can denote the best response of follower $i$ by the number of stakes it invest to the pool $p^*_i$. Then, the best response $p^*_i$ of follower $i$ can be expressed as a function of the pool's cost and fee as follows 
	\begin{equation}
	\label{bestresponse}
	p^*_i(\alpha,c) =
	\begin{cases}
	0 & \text{if $C_i< \dfrac{B_i\alpha R}{\sigma+\sum_{j=1}^{N}B_j}+c{\rm e}^{-B_i}$},\\
	B_i & \text{if $C_i\geq \dfrac{B_i\alpha R}{\sigma+\sum_{j=1}^{N}B_j}+c{\rm e}^{-B_i}$}.\\
	\end{cases}       
	\end{equation}
	\begin{theorem}
		Given a strategy of the leader, there exists an optimal strategy for every follower and this strategy is unique. 
	\end{theorem}
	\begin{IEEEproof}
		From (\ref{bestresponse}), it can be seen that for every fixed strategy of the leader, a unique best response of every follower can be straightforwardly determined.
	\end{IEEEproof}
	\subsubsection{Leader strategy}The backward induction mechanism~\cite{Game} can be used to find the best strategy of the leader, which is the strategy that yields the highest payoff given the best responses of all followers, i.e., we have
	\begin{equation}
	\label{eq:optimization}
	\begin{split}
	s^*_p&= \argmax_{s_p=(c,\alpha)} U_p(s_p,p^*_i)=\dfrac{\sigma}{\sigma+\sum_{j=1}^{N}B_j}R+\\&\sum_{i \in \mathcal{N}_p}\bigg( \dfrac{p^*_i\alpha}{\sigma+\sum_{j=1}^{N}B_j}R+c{\rm e}^{-B_i}\bigg).
	\end{split}
	\end{equation}
	
	Since the total network stakes can be considered a constant, the profit from the pool owner's stake is also a constant (the first term in (\ref{eq:optimization})) and does not need to be optimized. Moreover, since $p^*_i(\alpha,c)$ can only take two values, i.e., 0 or $B_i$, it can be represented by a binary decision variable $x_i \in \mathbf{x}= \{x_1,\ldots,x_N\}$, such that when $x_i=1$, $p^*_i=B_i$ and when $x_i=0$, $p^*_i=0$. This helps to transform the optimization problem (\ref{eq:optimization}) into a Mixed-Integer Programming (MIP) optimization as follows:
	\begin{equation}
	\begin{aligned}
	\label{MIP}
	\max_{\alpha,c,\mathbf{x}} \quad & \sum_{i=1}^{N} x_i\bigg(\dfrac{B_i R \alpha}{\sigma +\sum_{j=1}^{N}	B_j}+c{\rm e}^{-B_i}\bigg),&\\
	\textrm{s.t.} \quad & \dfrac{B_i R \alpha}{\sigma +\sum_{j=1}^{N}B_j}+c{\rm e}^{-B_i}\leq L(1-x_i)+ C_i  &\forall i \in \mathcal{N},\\
	&x_i \in \{0,1\} &  \forall i \in \mathcal{N},   \\
	\end{aligned}
	\end{equation}
	where $L$ is a sufficiently large number. The goal of (\ref{MIP}) is to find the optimal values of $(\alpha,c,\mathbf{x})$ to maximize the pool's profit. The objective function represents the profit of the pool, where the stake pool can only charge the stakeholders who have invested in the pool. The first set of constraints ensures that only when the pool charges follower $i$ less than $C_i$, $x_i$ can take the value of 1, and thus the profit can be added to the total profit of the pool. The second set of constraints ensures that every $x_i$ is a binary number. However, the objective function is nonlinear, i.e., it contains a multiplication of two decision variables $x_i$ and $\alpha$, which makes it much more complex to solve~\cite{MILP}. Thus, we transform (\ref{MIP}) into an equivalent Mixed-Integer Linear Programming (MILP) model as follows:
	\begin{equation}
	\begin{aligned}
	\label{MILP}
	\max_{\alpha,c,\mathbf{x},\mathbf{y}} \quad & \sum_{i=1}^{N} y_i,\\
	\textrm{s.t.} \quad & \dfrac{B_i R \alpha}{\sum_{j=1}^{N}B_j}+c{\rm e}^{-B_i}   \leq L(1-x_i)+ C_i& \forall i \in \mathcal{N},\\
	&y_i-Lx_i \leq 0  & \forall i \in \mathcal{N},   \\
	&y_i-L(1-x_i) \leq \dfrac{B_i R \alpha}{\sum_{j=1}^{N}B_j}+c{\rm e}^{-B_i}  & \forall i \in \mathcal{N},   \\
	&x_i \in \{0,1\}  & \forall i \in \mathcal{N},   \\
	&y_i \in \mathbb{R}^+  & \forall i \in \mathcal{N}.   \\
	\end{aligned}
	\end{equation}
	
	The transformation from (\ref{MIP}) to (\ref{MILP}) is done by a standard transformation technique which ensures the equivalence of the two models~\cite{transform}. In particular, we introduce a new set of continuous variables $\mathbf{y}= \{y_1,\ldots,y_N\}$ which represents the profit which the pool can yield from follower $i$. Two new sets of auxiliary constraints, i.e., the second and third sets of constraints, are added to set the upper bound for $y_i$. If $x_i=0$, i.e., follower $i$ does not invest stakes to the pool, $y_i$ will be upper-bounded by $0$. If $x_i=1$, $y_i$ will be upper-bounded by $\dfrac{B_i R \alpha}{\sum_{j=1}^{N}B_j}+c{\rm e}^{-B_i}$. Thus, the optimal solution of (\ref{MILP}) consists of two optimal values of $\alpha$ and $c$ as shown in (\ref{eq:optimization}).
	
	\subsubsection{Existence of the Stackelberg equilibrium}The existence of the Stackelberg equilibrium is proven via the existence of the optimal solutions of (\ref{MILP}) in the following Theorem.
	\begin{theorem}
		There exists at least one Stackelberg equilibrium in the considered stake pool game.
	\end{theorem}
	\begin{IEEEproof}
		We prove that there exists at least one solution of (\ref{MILP}). This means that there exists at least one leader's optimal strategy. Since only the decision variable $x_i$ is a binary number in (\ref{MILP}), if we fix the value of $x_i, \forall i \in \mathcal{N}$, (\ref{MILP}) becomes a Linear Programming (LP) problem. By fixing the value of $x_i$, we can decompose (\ref{MILP}) into $2^N$ LP problems (there are $2^N$ different combinations of $x_i$'s values). Each LP problem has the form of (\ref{MILP}), except that all $x_i$ are constants instead of decision variables. In the LP problem where $\sum_{i}^{N} x_i=0$, the optimal objective value is 0. In each of the remaining LP problems, the feasible region is constrained by
		\begin{equation}
		\dfrac{B_i R \alpha}{\sum_{j=1}^{N}B_j}+c{\rm e}^{-B_i}=C_i, \forall i \in \mathcal{N}_p.
		\end{equation} 
		Since $\alpha,c \geq 0$, the feasible region is bounded as illustrated in Fig.~\ref{Fig:region}. As a result, each of these $2^N$ LP problems has at least one optimal solution~\cite{MILP}. 
		\begin{figure}[ht]
			\includegraphics[width=.8\columnwidth]{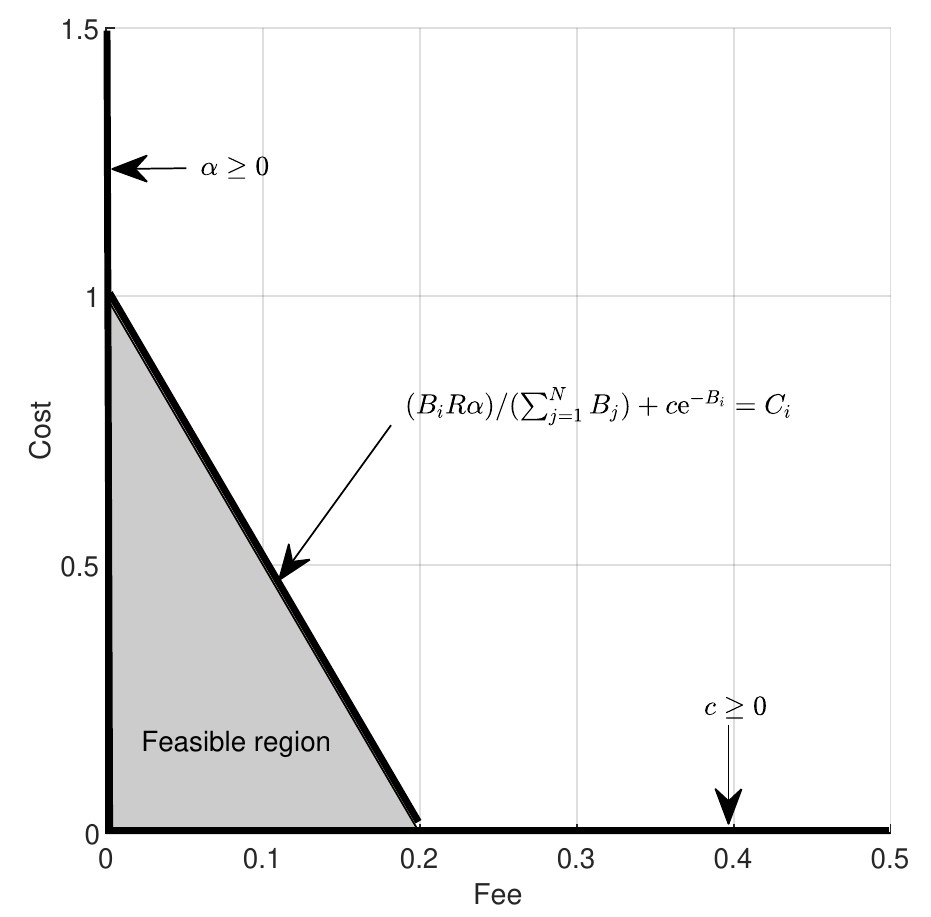}
			\centering
			\caption{An illustration of a bounded LP's feasible region.}
			\label{Fig:region}
		\end{figure}
		
		Since these $2^N$ LP problems enumerate all possible combinations of $x_i$ and each of these LP has at least one optimal solution, there exists at least one optimal solution of the MILP. Moreover, the existence of the best response of every follower is proven in Theorem 4. Therefore, there exists at least one Stackelberg equilibrium $(U^*_p,U^*_i)$ with the corresponding Stackelberg solution $(s^*_p,s^*_i)$ in this game.
	\end{IEEEproof}
	\subsubsection{Uniqueness of the Stackelberg equilibrium} Although there always exists at least one Stackelberg equilibrium in this game, the uniqueness of the equilibrium cannot be guaranteed because both $\alpha$ and $c$ are continuous variables. Consequently, there may be multiple pairs of $\alpha$ and $c$ to achieve the same optimal utility as will be shown later in Section~\ref{sec:simu}. In the conventional Stackelberg game model, the leader has only one primary priority, that is, to maximize the profit. Therefore, we propose a secondary priority for the leader, which is to minimize $\alpha$. This serves two purposes, i.e., to attract followers with high stakes (as the amount the pool charges via the fee is proportional to the stakes) and to determine the unique optimal strategy for the game (i.e., the unique optimal strategy for both the leader and followers). Under the proposed approach, we can always obtain the unique Stackelberg equilibrium that has the lowest fee among the Stackelberg equilibria.
	\section{Performance Evaluation}
	\label{sec:simu}
	\subsection{Parameter Settings}
	We first study three small game instances, i.e., $\mathcal{G}_1$ to $\mathcal{G}_3$, to clearly show the relation between the leader and the followers in different situations. In $\mathcal{G}_1$, we consider a small game consisting one stakeholder and one stake pool with $C_1=0.1$, $b_1=5$, $R=10$, and $\sigma=10$. Then, we extend this game to $\mathcal{G}_2$ by considering five followers with the same configurations as that of the follower in $\mathcal{G}_1$, while other parameters are unchanged. After that, we consider game $\mathcal{G}_3$. Parameters are similar as those of $\mathcal{G}_2$ except that the followers have different budgets $\mathbf{B} = (5,10,13,6,8)$, operational costs $\mathbf{C} = (0.1,0.3,0.2,0.6,0.5)$, and $R=50$. 
	
	To evaluate more general cases, we simulate 13 instances $\mathcal{G}_4$ to $\mathcal{G}_{16}$, each with 1,000 followers and different parameters as shown in Table~\ref{tab:instances}. Among them, the first five games $\mathcal{G}_4$ to $\mathcal{G}_8$ are simulated with network parameters, such as $R$, $\mathbf{C}$, and $\mathbf{B}$, generated based on several real-world PoS-based blockchain networks~\cite{Cardano,Algorand2,Cosmos,Tezos,Nem}. The follower's stakes and operational costs are generated randomly with normal distribution in the ranges listed in Table~\ref{tab:instances}. The results, including the optimal leader strategy, optimal profit, and percentage of the network stakes invested in the pool, are obtained by solving the MILP optimization (\ref{MILP}).
	\begin{table*}[]
		\centering
		\caption{Parameters and results of 13 simulation instances.}\label{tab:instances}
		\begin{tabular}{|c||c|c|c|c|c||c|c|c|c|c|c|}
			\hline
			\multirow{2}{*}{$\mathbfcal{G}$} & \multicolumn{5}{c||}{\textbf{Parameters}} &\multicolumn{4}{c|}{\textbf{Stackelberg equilibrium}}   \\ \cline{2-10} 
			& R                   &   $\mathbf{B}$ range                 & $\mathbf{C}$ range  & $\sigma$&Based on&$c^*$  & $\alpha^*$(\%)  & $U^*_p$  & \% stake of the pool   \\ \hline \hline
			4                    &1000 & [1,250] & [0.05,0.1]&1000&Cardano~\cite{Cardano}&3.2&4.0&28.95& 69.5
			\\ \hline
			5  & 200 & [1,1000] & [0.0001,0.15]&1000&Algorand~\cite{Algorand2}&0.06&1.6&1.81& 56.6
			\\ \hline
			6                      & 3.81 & [1,400] & [0.0001,0.002]&1000&Cosmos~\cite{Cosmos}&0.1&14.4&0.35& 61.2
			\\ \hline		 
			7         & 78 & [80,160] & [0.0001,0.02]&1000&Tezos~\cite{Tezos}&40.1&6.1&2.29& 48.9
			\\ \hline		
			8                    & 500 & [1,5000] & [0.001,0.3]&1000&NEM~\cite{Nem}&0.003&13.01&40.92& 62.9
			\\ \hline \hline	
			9                    &\textbf{100} & [1,250] & [0.05,0.1]&1000&Cardano &0.003&40.4&28.08& 69.5
			\\ \hline
			10                    &\textbf{10000} & [1,250] & [0.05,0.1]&1000&Cardano&0.207&0.4&29.13& 69.5
			\\ \hline \hline	
			11                    &1000 & [1,250]&\textbf{[0.01,0.02]}&1000&Cardano&0.04&0.8&5.82&69.5
			\\ \hline  
			12                    &1000 & [1,250] & \textbf{[0.25,0.5]}&1000&Cardano&0.04&20.5&140.54& 69.5
			\\ \hline	\hline 
			13                    &1000 & \textbf{[1,25]} & [0.05,0.1]&1000&Cardano&0.2&4.7&36.51& 72.1
			\\ \hline
			14                   &1000 & \textbf{[1,2500]} & [0.05,0.1]&1000&Cardano&356.1&4.0&28.21& 70.1
			\\ \hline \hline
			15                   &1000 & [1,250] & [0.05,0.1]&\textbf{1}&Cardano&0.04&4.0&28.31& 69.5
			\\ \hline
			16                   &1000 & [1,250] & [0.05,0.1]&\textbf{100000}&Cardano&0.02&10.9&28.15& 69.5
			\\ \hline
		\end{tabular}
	\end{table*}
	\subsection{Numerical Results}
	\subsubsection{Small Cases}
\begin{figure}
		\begin{subfigure}{\columnwidth}
			\centering
			\includegraphics[width=\linewidth]{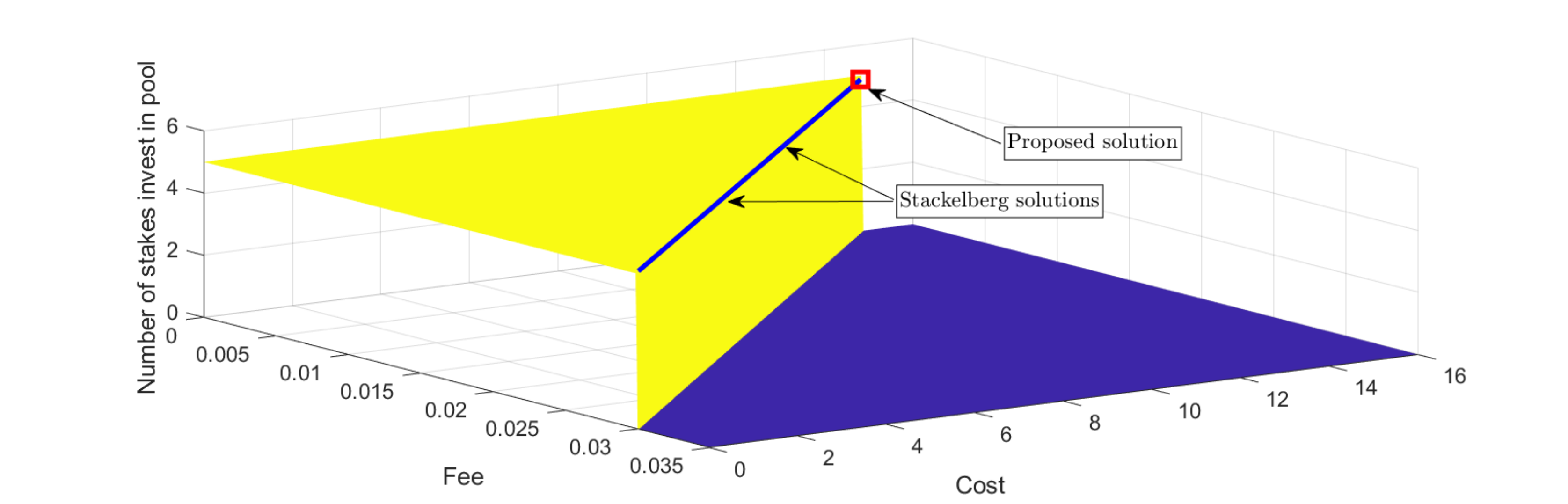}
			\caption{Best response function of follower 1}
			\label{Fig:fig3a}
		\end{subfigure}\\
		\begin{subfigure}{\columnwidth}
			\centering
			\includegraphics[width=\linewidth]{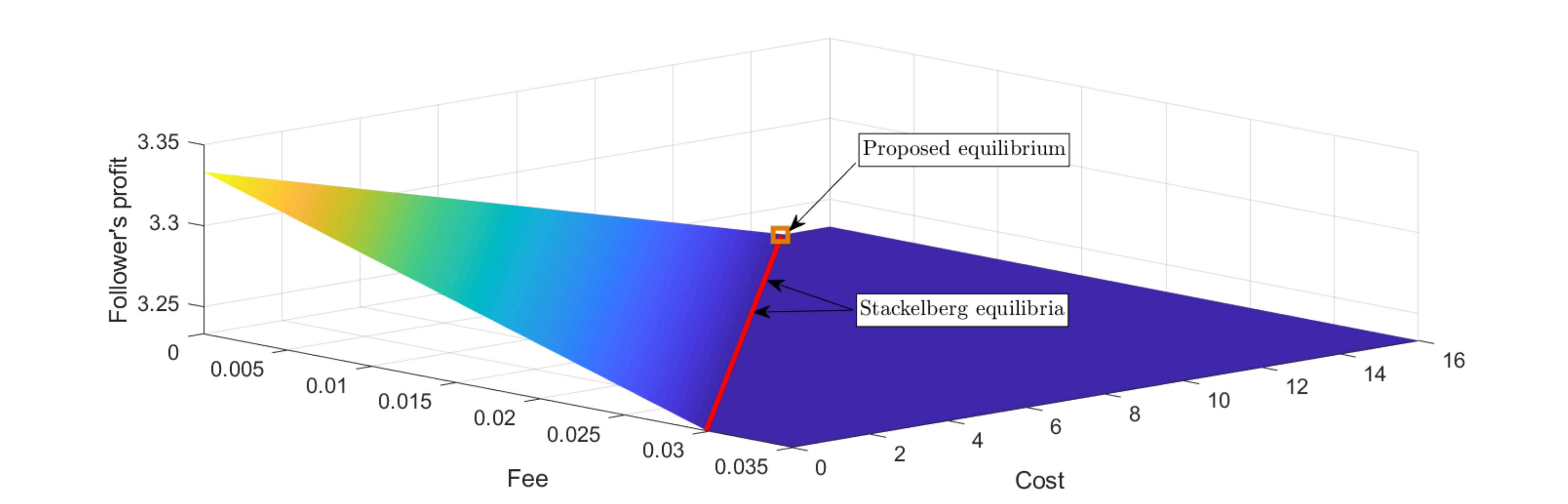}
			\caption{Profit of follower 1}
			\label{Fig:fprofitg1}
		\end{subfigure}\\
		\begin{subfigure}{\columnwidth}
			\centering
			\includegraphics[width=\linewidth]{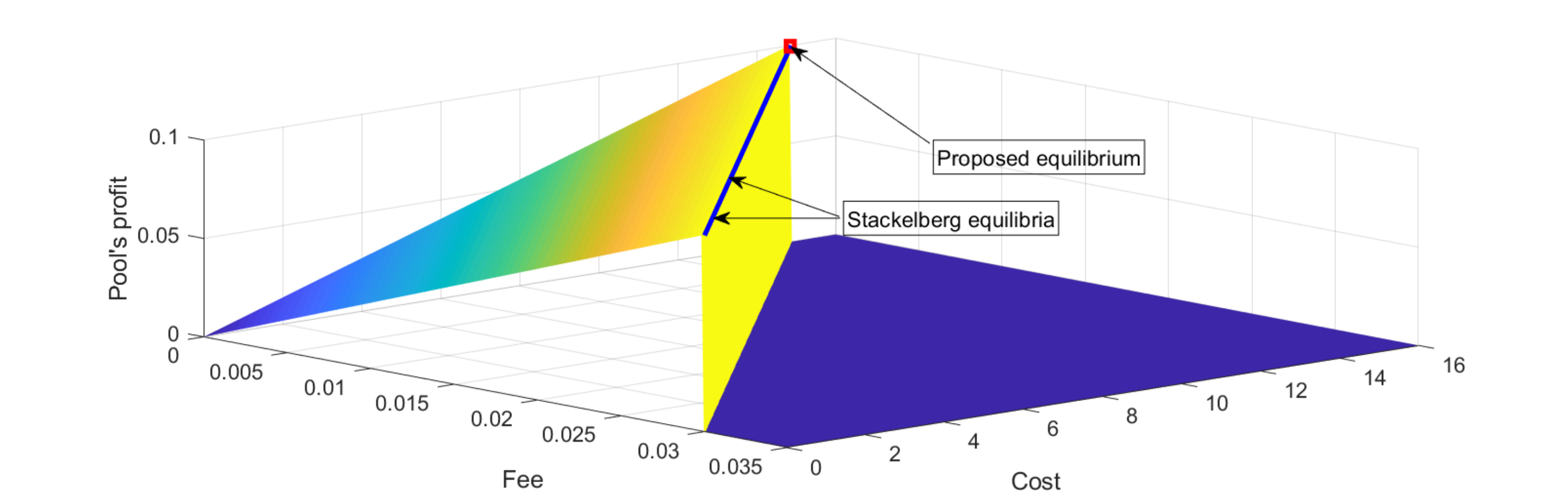}
			\caption{Pool's profit}
			\label{Fig:fig4}
		\end{subfigure}
		\caption{Profit and best response of the leader and follower in $\mathcal{G}_1$.}
		\label{Fig:g1}
	\end{figure}

	The best response function of follower 1 in $\mathcal{G}_1$ is illustrated in Fig.~\ref{Fig:g1}\subref{Fig:fig3a}. Based on its best response, the profit of follower 1 can be determined. In this game, the profit of the follower decreases as the pool's fee and cost increase as shown in Fig.~\ref{Fig:g1}\subref{Fig:fprofitg1}, but it is still higher than self-mining. The profit of the pool is illustrated in Fig.~\ref{Fig:fig4}. Since there is only one follower in $\mathcal{G}_1$, the profit of the pool only comes from follower 1, and thus it is upper-bounded by $C_1$. In this game, any pair of ($c,\alpha$) that satisfies $\dfrac{\alpha R B_i}{\sigma+C_i}+c{\rm e}^{-B_i}=C_i=\dfrac{50}{15}\alpha+0.007c=0.1$ is a Stackelberg solution, which leads to multiple Stackelberg equilibria. Nevertheless, under our proposed approach, we can find the unique Stackelberg equilibrium for this game at $(c^*,\alpha^*)=(14.8,0)$.
	
\begin{figure}
		\begin{subfigure}{\columnwidth}
			\centering
			\includegraphics[width=\linewidth]{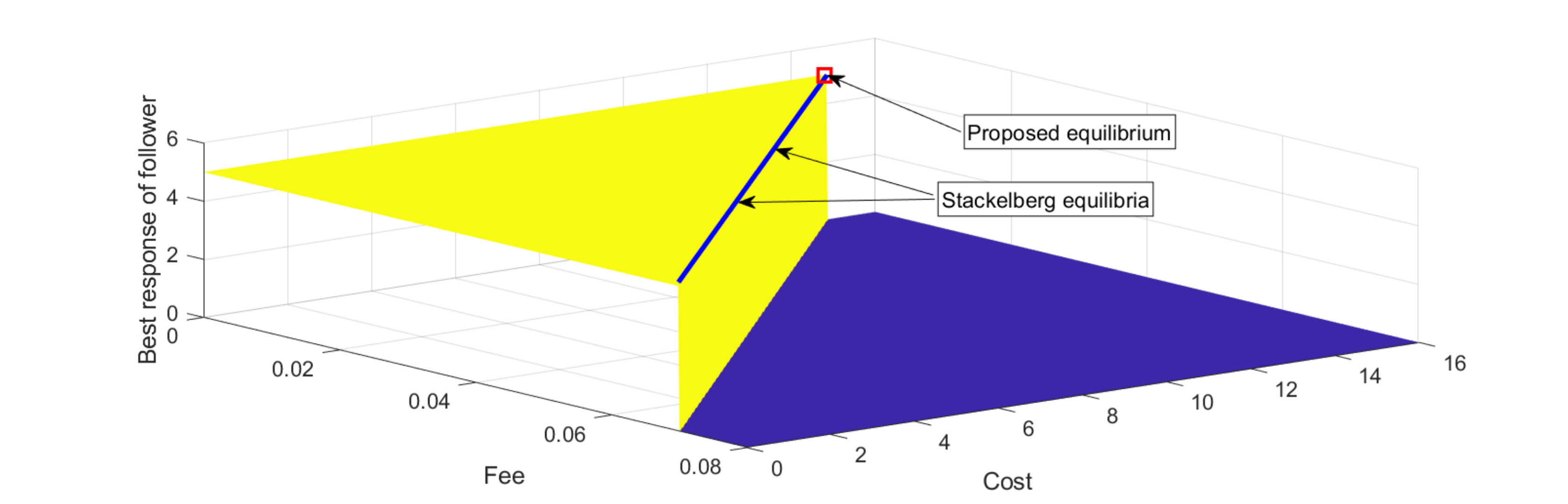}
			\caption{Best response function of follower 1}
			\label{Fig:bestresponse_G2}
		\end{subfigure}\\
		\begin{subfigure}{\columnwidth}
			\centering
			\includegraphics[width=\linewidth]{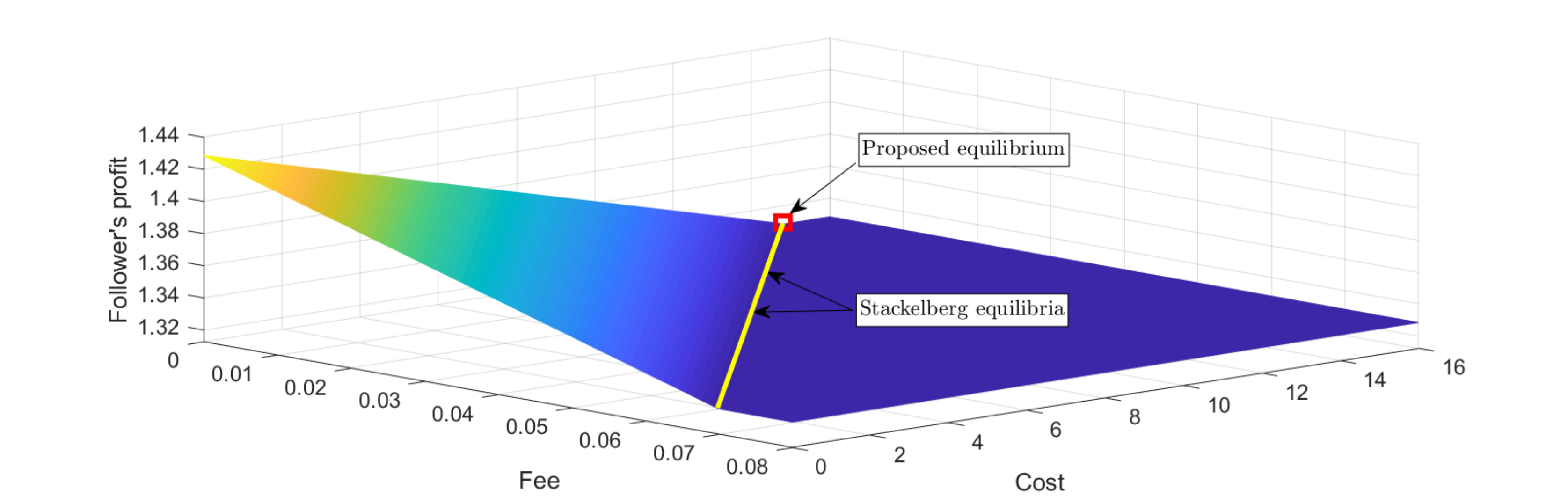}
			\caption{Profit of follower 1}
			\label{Fig:fprofitg2}
		\end{subfigure}\\
		\begin{subfigure}{\columnwidth}
			\centering
			\includegraphics[width=\linewidth]{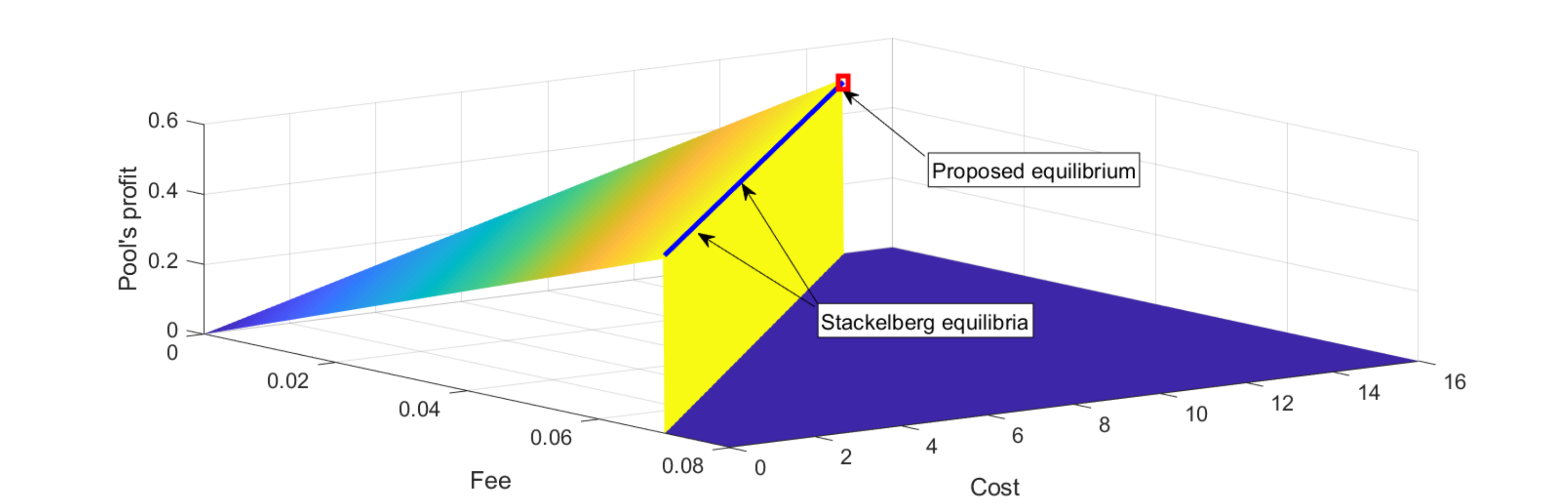}
			\caption{Pool's profit}
			\label{Fig:fig5}
		\end{subfigure}
		\caption{Profit and best response of the leader and follower in $\mathcal{G}_2$.}
		\label{Fig:g2}
	\end{figure}

	In $\mathcal{G}_2$, since the followers have the same budgets and operational costs, their best response and profit functions are the same, which are illustrated in Fig.~\ref{Fig:bestresponse_G2} and Fig.~\ref{Fig:fprofitg2}, respectively. These functions are similar to that of $\mathcal{G}_1$, except that the fee threshold is higher (7\%). This is because there are more followers in $\mathcal{G}_2$, and thus ($c,\alpha$) must satisfy $\dfrac{\alpha R B_i}{\sigma+C_i}+c{\rm e}^{-B_i}=C_i=\dfrac{50}{35}\alpha+0.007c=0.1$.
	The pool's profit in $\mathcal{G}_2$ is illustrated in Fig.~\ref{Fig:fig5}, which is upper-bounded by $5C_i$ in this game. The unique proposed equilibrium of this game has a corresponding solution $(c^*,\alpha^*)=(14.8,0)$ as shown in Fig.~\ref{Fig:fig5}. 
	
\begin{figure}
		\begin{subfigure}{\columnwidth}
			\centering
			\includegraphics[width=\linewidth]{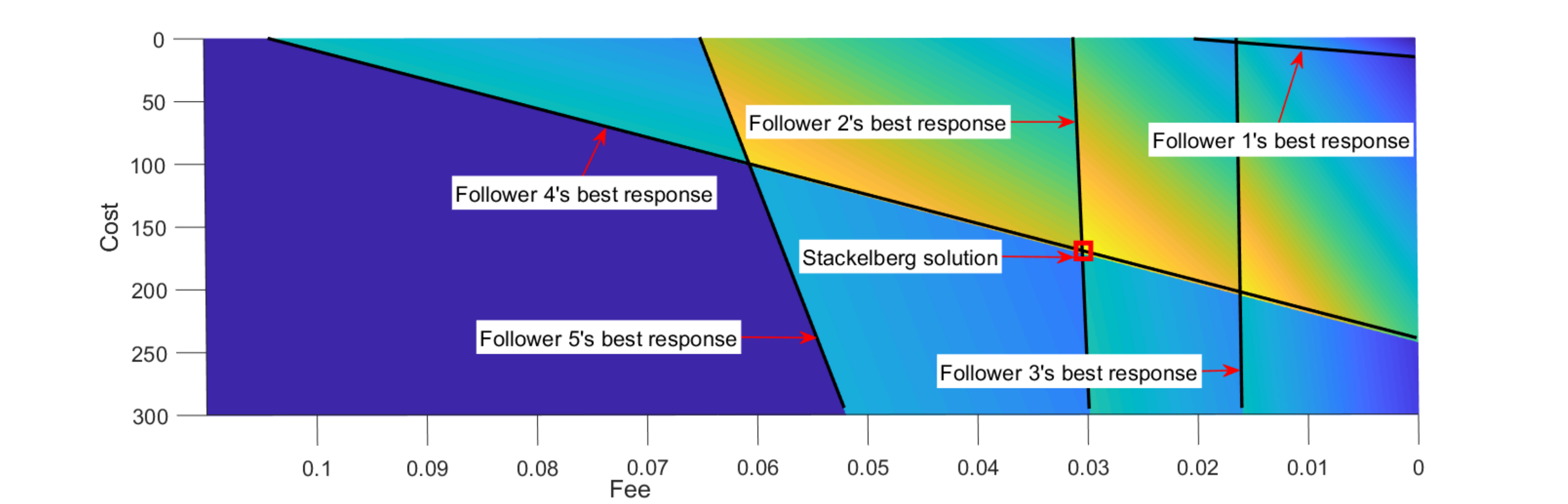}
			\caption{Best responses of followers}
			\label{Fig:bestresponse_G3}
		\end{subfigure}\\
		\begin{subfigure}{\columnwidth}
			\centering
			\includegraphics[width=\linewidth]{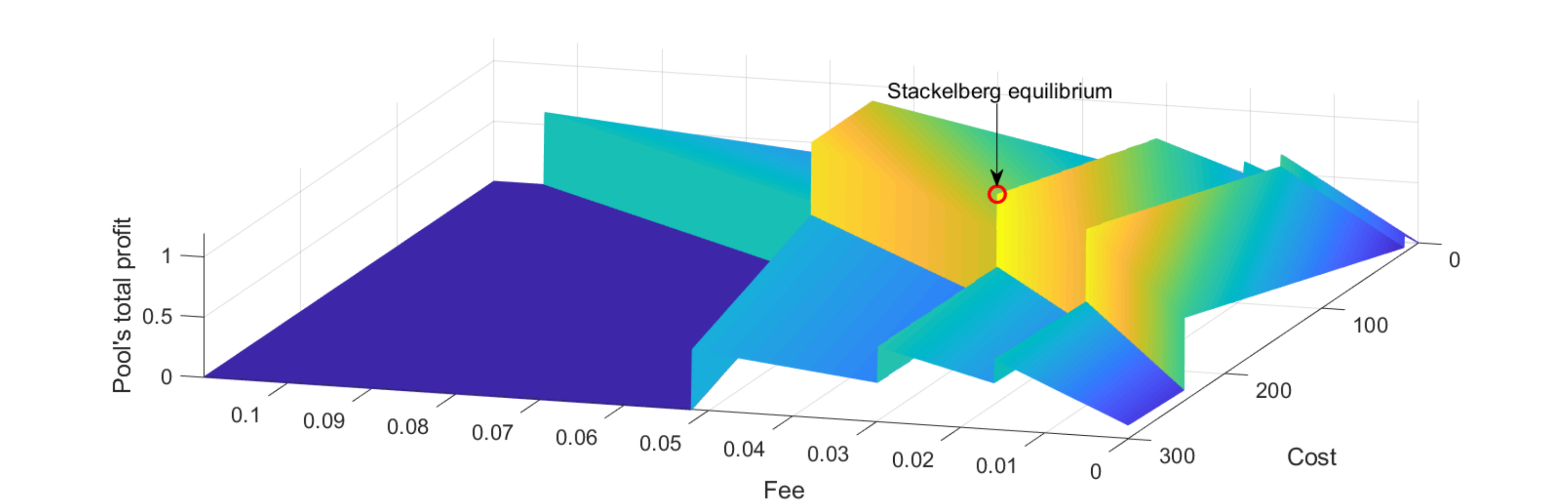}
			\caption{Pool's total profit}
			\label{Fig:fig6}
		\end{subfigure}\\
		\begin{subfigure}{\columnwidth}
			\centering
			\includegraphics[width=\linewidth]{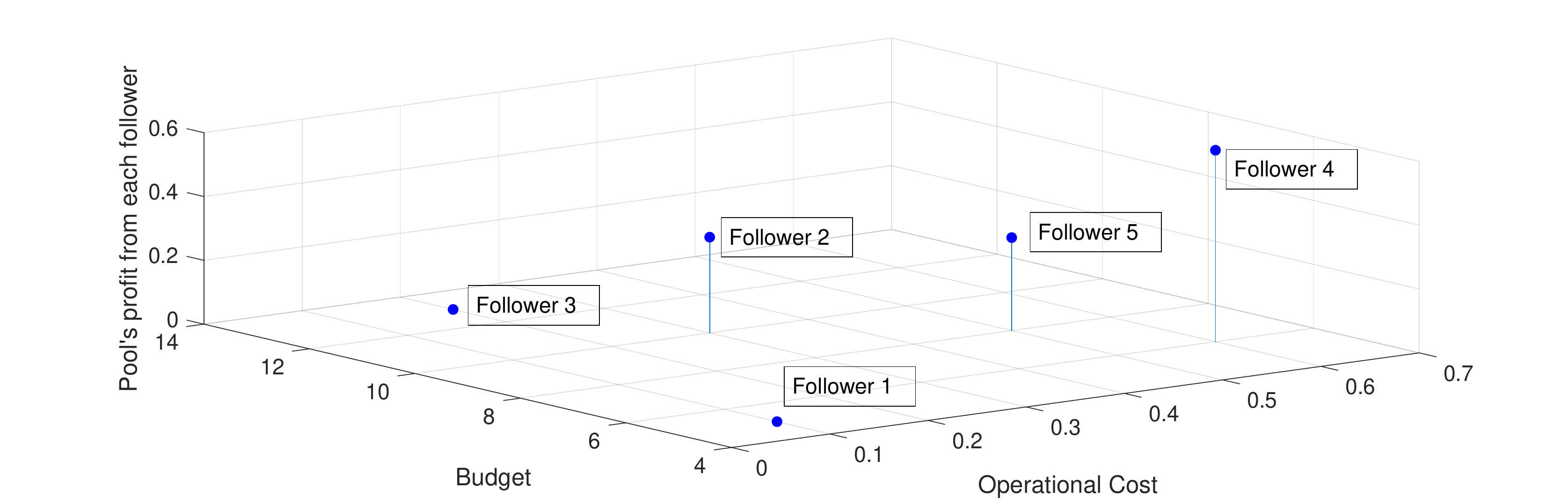}
			\caption{Pool's profit from each follower}
			\label{Fig:pooliniG3}
		\end{subfigure}
		\caption{Profit and best response of the leader and followers in $\mathcal{G}_3$.}
		\label{Fig:g3}
	\end{figure}

	In $\mathcal{G}_3$, each follower's best response is illustrated in~\ref{Fig:bestresponse_G3}. Typically, the higher a follower's budget is, the higher cost and the lower fee that follower is willing to accept, and vice versa. For example, follower 3 with the highest budget only accepts a fee of no more than $1.6\%$, and follower 1 with the lowest budget only accepts a cost lower than 15. This is because the budget is proportional to the fee the pool charges, while the cost decreases exponentially as the budget increases. The pool's profit in $\mathcal{G}_3$ is illustrated in Fig.~\ref{Fig:fig6}, with the leader's optimal strategy $(c^*,\alpha^*)=(171.3,3.0\%)$ and optimal profit $U^*_p=1.19$. Fig.~\ref{Fig:pooliniG3} illustrates the profit the pool receives from each follower. Interestingly, at the obtained Stackelberg equilibrium of $\mathcal{G}_3$, the follower with the highest stake, i.e, follower 3, does not invest to the pool. The reason is that follower 3 has a relatively low operational cost, and thus the follower is more inclined to mine if the pool's cost and fee are too high. If the pool tries to incentivize all followers to invest by reducing $\alpha$ and $c$, its profit is only $U_p=0.68$.
	\subsubsection{General Cases} 
\begin{figure}[ht]
		\includegraphics[width=\columnwidth]{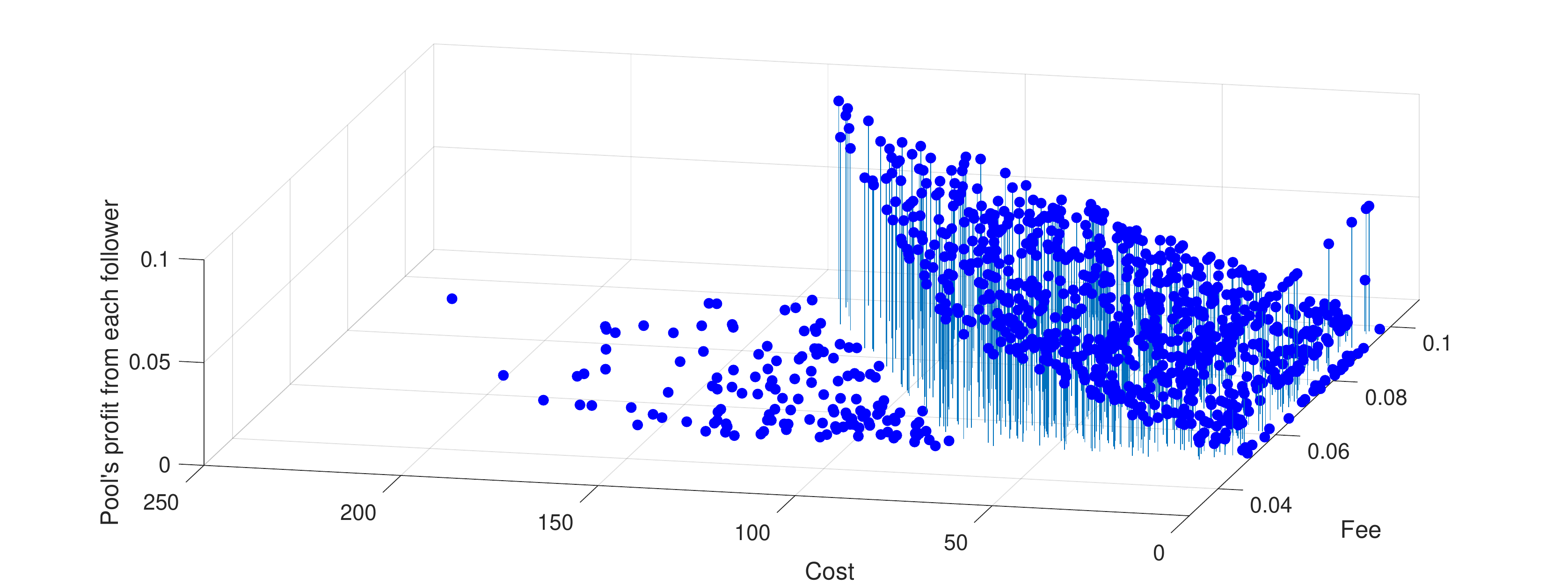}
		\centering
		\caption{Pool's profit from each follower in $\mathcal{G}_4$.}
		\label{Fig:fig7}
	\end{figure}

The results of more general cases are shown in Table~\ref{tab:instances}. The five instances $\mathcal{G}_4$ to $\mathcal{G}_8$ are simulated with parameters adopted from several real-world blockchain networks~\cite{Cardano,Algorand2,Cosmos,Tezos,Nem}. The results show that the leader's optimal strategy and profit are significantly influenced by the network's parameters. For example, we obtain the optimal solution of $\mathcal{G}_4$ where $(c^*,\alpha^*)=(3.2,4.0\%)$, $U^*_p=28.95$, and approximately 69.5\% of the total network's stakes (including $\sigma$) are invested to the pool. The profit that the pool earns from each follower depends on each follower's budget and operational cost, as shown in Fig.~\ref{Fig:fig7}. Typically, a follower with higher cost and budget can give the pool more profit. However, similar to $\mathcal{G}_3$, if the budget is too high, the follower might not want to invest stakes to the pool, e.g., the followers with budget $B_i$ greater than 150 do not join the pool in $\mathcal{G}_4$. 
	
	\subsubsection{Impacts of Parameters}The last eight games $\mathcal{G}_9$ to $\mathcal{G}_{16}$ are simulated to study the impacts of important parameters $R$, $\mathbf{B}$, $\mathbf{c}$, and $\sigma$, on the game's outcome. The impacts of those parameters are briefly described as follows:
	\begin{itemize}
		\item \textit{Block reward $R$:} $\mathcal{G}_9$ and $\mathcal{G}_{10}$ are simulated to show the impact of $R$. As $R$ increases, the pool's profit increases. However, the followers' operational costs are constant. Therefore, the pool has to decrease $\alpha$ when $R$ increases, otherwise the followers will self-mine.
		\item \textit{Operational costs $\mathbf{C}$:} $\mathcal{G}_{11}$ and $\mathcal{G}_{12}$ show how the followers' operational cost impacts the game's outcome. As the $\mathbf{C}$ increase, the pool can increase its profit by increasing $\alpha$. The reason is that the followers' profits from self-mining are inversely proportional to the $\mathbf{C}$, and thus self-mining becomes less profitable if $\mathbf{C}$ are too high.
		\item \textit{Budgets $\mathbf{B}$:}  $\mathcal{G}_{13}$ and $\mathcal{G}_{14}$ show that as the budgets of followers increase, the pool can increase $c$ but it has to reduce $\alpha$. This is because the profit the pool receives via $\alpha$ is proportional to $\mathbf{B}$, while the profit the pool gets from $c$ decreases exponentially as $\mathbf{B}$ increase. Moreover, as $\mathbf{B}$ increase, the stakeholders invest fewer stakes to the pool and consequently the pool's profit decreases. The reason is that when $\mathbf{B}$ increase, the profit from self-mining also increases, and thus the followers prefer to self-mine.
		\item \textit{The pool owner's stake $\sigma$:} The last two games show that as $\sigma$ increases, although there are more stakes invested in the pool, its profit slightly decreases. The reason is that $\sigma$ is inversely proportional to the pool's profit from each follower, and thus increasing $\sigma$ means that the pool charges less from each follower. Consequently, the pool's profit decreases even though more followers invest to the pool. 
	\end{itemize}
	\subsection{Summary of Findings}
	The key findings of the considered stake pool game are summarized as follows:
	\begin{itemize}
		\item We have proved that for a rational stakeholder, its best strategy is to invest all stakes from its budget to the blockchain network.
		\item  We have proved that for each stakeholder, its best strategy is to invest all its stakes either to the pool or for self-mining.
		\item We have proposed an approach for the leader to decide its optimal strategy. Under this approach, there always exists the optimal and unique best strategies for the stakeholders and the stake pool owner. This approach also helps the stake pool to attract stakeholders with high stakes.
	\end{itemize}
	%----------------------------
	%----------------------------
	
	%++++++++++++++++++++++++++++++++++++++++++++++
	%++++++++++++++++++++++++++++++++++++++++++++++
	
	\section{Conclusion}
	\label{sec:Sum}
	To address the problem of roaming fraud for mobile service providers, we have proposed BlockRoam, a novel blockchain-based roaming management system which consists of our thoroughly analyzed PoS consensus mechanism and a smart-contract-enabled roaming management platform. Moreover, we have analyzed and showed that BlockRoam's security and performance can be enhanced by incentivizing more users to participate in the network. Therefore, we have developed an economic model based on Stackelberg game to jointly maximize the profits of network users, thereby incentivizing their participation. We have analyzed and determined the best strategies for the stakeholders and the stake pool. We have also proposed an effective solution that results in a unique equilibrium for our economic model. Lastly, we have evaluated the impacts of important parameters on the strategies and the equilibrium of the game. The proposed economic model can help the mobile service providers to earn additional profits, attract more investment to the blockchain network, and enhance the network's security and performance.
	
	\ifCLASSOPTIONcaptionsoff
	\newpage
	\fi

	\vspace{0.7cm}
	%==============================================================
	%==============================================================

\end{document}